# Leveraging Sequence Purification for Accurate Prediction of Multiple Conformational States with AlphaFold2

Enming Xing[a], Junjie Zhang[a], Shen Wang[a], Xiaolin Cheng[a,b*]

[a] *Division of Medicinal Chemistry and Pharmacognosy, College of Pharmacy, The Ohio State University, Columbus OH, 43210, USA*

[b] *Translational Data Analytics Institute, The Ohio State University, Columbus, OH 43210, USA*

* *Corresponding Authors: Xiaolin Cheng (cheng.1302@osu.edu)*

## Abstract

AlphaFold2 (AF2) has transformed protein structure prediction by harnessing co-evolutionary constraints embedded in multiple sequence alignments (MSAs). MSAs not only encode static structural information, but also hold critical details about protein dynamics, which underpin biological functions. However, these subtle co-evolutionary signatures, which dictate conformational state preferences, are often obscured by noise within MSA data and thus remain challenging to decipher. Here, we introduce AF-ClaSeq, a systematic framework that isolates these co-evolutionary signals through sequence purification and iterative enrichment. By extracting sequence subsets that preferentially encode distinct structural states, AF-ClaSeq enables high-confidence predictions of alternative conformations. Our findings reveal that the successful sampling of alternative states depends not on MSA depth but on sequence purity. Intriguingly, purified sequences encoding specific structural states are distributed across phylogenetic clades and superfamilies, rather than confined to specific lineages. Expanding upon AF2's transformative capabilities, AF-ClaSeq provides a powerful approach for uncovering hidden structural plasticity, advancing allosteric protein and drug design, and facilitating dynamics-based protein function annotation.

**Main:**

AlphaFold2 (AF2) has transformed protein structure prediction[1-3], achieving unprecedented accuracy by integrating deep learning with co-evolutionary constraints[4-6] extracted from Multiple Sequence Alignments (MSAs). In the early 2000s, Ranganathan and colleagues developed Statistical Coupling Analysis (SCA) to identify co-evolving residues within protein families, quantifying evolutionary dependencies between amino acid positions through sequence alignments[7, 8]. These co-evolving residues often correspond to native contacts essential for maintaining protein stability, conformation and function. More recently, methods such as EVcoupling utilized co-evolutionary information from MSAs to predict residue-residue contacts, enabling accurate predictions of protein structures and protein-protein interactions[6, 9-14]. By integrating MSA of a query sequence into deep neural networks, AF2 significantly improves prediction accuracy, even for the most challenging protein structures.

Proteins are inherently dynamic, often interconverting between multiple conformational states crucial for their biological functions[15-17]. However, AF2 typically predicts a single structure for a given sequence[18, 19], limiting its utility for proteins that adopt distinct conformations under different conditions. To address this limitation, various strategies have been developed to enable AF2 to predict alternative conformations[20-27]. Wayment-Steele et al. introduced a sequence-based clustering method that successfully predicted multiple states for proteins with known structural heterogeneity, such as KaiB and RfaH[28]. Another promising approach involves reducing MSA depth through stochastic subsampling or representative subset selection, which introduces variability and facilitates alternative conformation prediction for transporters and G-protein-coupled receptors (GPCRs)[24, 29]. Additionally, methods such as MSA column masking and mutagenesis, exemplified by tools like SPEACH_AF[25], AFsample[21], and AFsample2[22], have been explored to bias AF2 predictions toward specific conformational states. Instead of manipulating MSAs, Bryant and Noé developed Cfold[26], a structural prediction network trained on conformationally split protein datasets. Although Cfold performed well for proteins undergoing moderate conformational changes, its predictive accuracy declined significantly for systems exhibiting large structural transitions (e.g., TM-score changes >0.4).

Proteins sharing a common ancestor follow divergent evolutionary trajectories, accruing mutations that drive structural and functional diversification[30-34]. Over time, homologous proteins diverge into distinct groups with conformational preferences tailored to their specific functions. Hence, the co-evolutionary signals embedded in an MSA do not correspond to a single structure but a mixture of conformational states. Morcos et al. employed direct coupling analysis to identify correlated mutations within MSAs, revealing coevolving residue pairs that do not necessarily co-exist in a single conformation[35]. Given this inherent heterogeneity, AF2 neural networks struggle to disentangle overlapping co-evolutionary signals associated with distinct structural states. Instead, these signals are averaged or obscured, confounding accurate structural prediction. As depicted in **Fig. 1**, when an MSA contains sequences encoding multiple conformational states, AF2 tends to predict an ensemble-averaged structure with low confidence due to frustration in the co-evolutionary constraints; it is impossible for a single structure to satisfy these conflicting constraints simultaneously. Conversely, when an MSA predominantly consists of sequences encoding a single conformation, or when structural variations are localized, predicting alternative conformations becomes inherently difficult, as AF2 always biases toward predicting the most frequently encoded state. Thus, the primary challenge for AF2 in predicting multiple conformations lies in the mixed nature of MSAs, which aggregate sequences with divergent conformational preferences. If MSAs could be systematically classified and refined into subsets that predominantly encode individual conformational states, AF2 could be directed to produce high-confidence predictions for distinct conformations.

However, deciphering conformational heterogeneity from MSAs remains a formidable challenge[8, 36], given the high noise and long-range correlations inherent in co-evolutionary data. Conventional clustering based on sequence similarity fails to reliably separate co-evolutionary patterns associated with distinct conformational states. While SCA can extract this information, it becomes intractable without prior structural data for at least two states. Recent work by Guan et al. has sought to address these challenges by incorporating energy landscape features into AF2[37]. However, this approach relies on predefined energy frustration scores, which may be insufficient to capture subtle frustration patterns in MSAs, particularly those linked to small-amplitude local conformational changes. Moreover, while an individual sequence may inherently favor a particular conformational state, it lacks sufficient co-evolutionary depth required for reliable structural predictions. Thus, an exhaustive exploration of MSA subsets appears necessary to assign individual sequences to specific conformations, rendering this approach computationally prohibitive due to the vast search space.

In this work, we introduce AF-ClaSeq, a sequence purification approach that enables AF2 to predict alternative conformational states by classifying sequences within MSA space. AF-ClaSeq employs bootstrapping and M-fold sampling to generate randomly populated MSA subsets, each used to predict a structure with AF2. A voting mechanism then assigns each sequence to the structural state most frequently observed across predictions for MSAs containing that sequence. This process generates purified MSA subsets enriched in sequences encoding distinct conformational states. AF-ClaSeq addresses challenges similar to those tackled by AlphaCryo4D[38], which classifies large pools of noisy 2D cryo-EM images into distinct 3D structural ensembles. Purified MSAs not only enhance AF2's ability to predict alternative states with high accuracy and confidence, but also provide insights into the interplay between sequence evolution, protein dynamics, and function. Remarkably, even small purified MSA subsets (e.g., containing as few as ten sequences) can yield high-confidence structural predictions with low variability, underscoring the power of sequence purification in uncovering hidden structural plasticity.

## Results

### Sequence purification via bootstrapping and voting mechanisms

The AF-ClaSeq pipeline comprises three key steps: (1) bootstrapping and M-fold sampling of sequences from an input multiple sequence alignment (MSA); (2) structural prediction using each sampled MSA subset; and (3) structure-guided sequence voting and recompilation. As illustrated in **Fig. 2**, the workflow begins with the generation of a MSA for a query sequence through the MMseqs2 server[39-44] or DeepMSA2[45, 46]. The MSA is divided into M+1 groups using a leave-one-group-out bootstrapping procedure, ensuring each sequence appears exactly M times across all subsets. Structure prediction is then performed on each MSA subset using ColabFold[44, 47], and the resulting structures are binned based on a reaction coordinate, producing a probability distribution of conformational states. A sequence voting mechanism tracks the structural preferences of individual sequences, assigning each to the bin where it most frequently appears. This enables the extraction of structure-specific sequences, which are subsequently recompiled into purified MSAs for distinct conformational states.

AF-ClaSeq enhances the contribution of individual sequences to ensemble structural predictions, effectively identifying structure-specific sequences from a large and noisy sequence pool. The approach was validated across eight diverse protein systems, capturing conformational dynamics spanning multiple scales, including loop dynamics, secondary structure rearrangements, hinge-bending motions, and metamorphic fold-switching[18, 19]. We present five representative systems below, with additional examples included in the **Extended Data Figs 1-4**.

**Predicting apo and ligand-bound states of adenylate kinase with purified sequences**

Adenylate kinase (AdK) plays a pivotal role in cellular energy homeostasis, catalyzing the reversible transfer of phosphate groups between adenine nucleotides. This function involves binding of ATP or ADP, triggering substantial conformational changes as the enzyme transitions from an open (apo) to a closed (ligand-bound) state[48-50]. Structures of the open (green) and closed (purple) states have been determined[48, 50], but structural predictions (grey) using AF2 with raw MSAs consistently yield closed-like conformations (**Fig. 3a**). Even when multiple predictions are generated from the same MSA, the results exhibit only minor variation and fail to capture the open state, consistent with previous studies[20, 51].

To overcome this limitation, we employed AF-ClaSeq to identify sequences capable of accurately predicting both states of AdK. Starting with 3,674 sequences, the MSA was shuffled and split into 368 groups of 10 sequences for 367 times to generate 135,056 MSA subsets. A single structural prediction was obtained from each MSA. Root mean square deviations (RMSDs) of the predicted structures relative to the experimental apo and ligand-bound states were computed and plotted in **Fig. 3b**. The RMSD values were further divided into 30 bins, which represent structural microstates spanning the transition between the two states (**Fig. 3c**). Sequence voting was then applied to identify the sequences contributing most significantly to each bin. Upon voting, sequences associated with the 9$^{th}$ bin (apo state panel) and the 8$^{th}$ bin (ligand-bound state panel) were designated as purified sequences for the apo and the ligand-bound states, respectively.

To evaluate the effectiveness of the AF-ClaSeq approach, structural predictions using purified sequences were compared with those generated using randomly selected sequences from the original MSA. For a fair comparison, the number of randomly selected sequences was matched to the size of each purified MSA subset. 40 structures were generated for each MSA using eight random seeds. Purified sequences consistently produced highly accurate predictions for both apo (RMSD: 1.0-1.5 Å) and ligand-bound (RMSD: 1.2-1.8 Å) states (**Fig. 3d**). The predicted models with the highest pLDDT scores aligned remarkably well with the corresponding experimental structures, with an RMSD of 1.27 Å for the apo state and 1.29 Å for the ligand-bound state (**Fig. 3e**). In contrast, same number of randomly selected sequences exhibited substantial variability (RMSD: 3.4-5.2 Å for apo and 2.2-3.4 Å for ligand-bound). Intriguingly, purified sequences were neither evolutionarily clustered nor characterized by uniform sequence coverage (**Extended Data Fig. 5**), underscoring the method's ability to identify structurally specific sequences independent of evolutionary proximity.

To minimize reliance on prior knowledge of end-state structures, we developed a universal method for defining structural states across predictions. A ResNet-based variational autoencoder (VAE) was used for dimensionality reduction of Cα atom coordinates, allowing the projection of all predicted AdK structures onto a 2D UMAP space (**Extended Data Fig. 6a**). Structural states were then assigned by segmenting this projection into discrete bins, followed by sequence voting (**Extended Data Fig. 6b**). Sequences corresponding to bins 3_8 and 18_13 were identified as purified MSAs for the apo and the ligand-bound states, respectively. These purified MSAs consistently yielded highly accurate predictions for both apo (RMSD: 2.0-2.3Å) and ligand-bound (RMSD: 1.6-1.8Å) conformations. Notably, densely populated regions in the UMAP corresponded to conformations along a transition pathway between apo and ligand-bound states (**Extended Data Figs. 6c, 6d**), shedding light on the structural dynamics underlying conformational transitions.

A previous study by Guan et al. introduced a frustration-filtering strategy that coupled AF2 with protein energy landscape analysis to predict alternative conformations and transition pathways of AdK[37]. By subsampling MSA sequences with different energy thresholds, this approach successfully predicted nearly closed (RMSD: 0.80 Å) and open (RMSD: 1.63 Å) structures. However, a key limitation of this approach is its dependence on accurate frustration score calculations, a highly nontrivial task akin to challenges associated with energy functions in molecular dynamics and molecular docking. Moreover, the strategy often requires prior structural knowledge to generate functionally relevant alternative structures. In contrast, our data-driven AF-ClaSeq platform employs a structure-guided sequence purification strategy that eliminates the need for energy-based calculations and requires minimal prior structural knowledge.

**Predicting active and inactive states of ABL1 with purified sequences**

The ABL1 kinase is a key player in cellular signaling, regulated by conformational changes that control its activity[52]. A critical step in ABL1 activation involves the transition of the activation loop (A-loop) between the "DFG-out" and "DFG-in" conformations[53]. In the "DFG-in" conformation, the aspartate residue coordinates a magnesium ion required for ATP binding, stabilizing the kinase in its active state. Conversely, the "DFG-out" conformation repositions the loop, disrupting the catalytic machinery and rendering the kinase inactive[54]. This structural plasticity allows ABL1 to switch between active and inactive states in response to cellular signals.

Experimental structures of both the active (DFG-in) and inactive (DFG-out) states have been resolved[53], with the A-loop adopting distinct conformations in each state (**Fig. 4a**). However, default AF2 predictions using raw MSAs consistently produce only the active DFG-in state, failing to capture the inactive conformation. A DeepMSA2-generated MSA for ABL1 yielded 49,839 sequences, but the sequences contributing to the inactive state constituted a minor fraction, rendering direct application of the voting-based M-fold sampling computationally prohibitive. To address this challenge, we developed an iterative enrichment strategy to construct a smaller, focused MSA subset for effective voting.

The enrichment process began by shuffling and splitting sequences into groups of 20, repeating this process 10 times, followed by structural prediction for each group (**Extended Data Fig. 7a**). RMSDs for the A-loop were calculated by aligning the predicted structures with the inactive state. In the first iteration, predicted structures exhibited RMSDs of 15-20 Å, indicating poor resemblance to the inactive conformation. We selected sequences corresponding to the lowest 20$^{th}$ percentile of RMSD values and compiled them into a refined MSA for the next iteration. This process was repeated for 16 iterations, progressively enriching the MSA containing sequences that contributed to lower RMSD values for the A-loop compared to the inactive state. Over successive iterations, the predicted structures increasingly resembled the inactive state. At the end of this process, we collected all structures with RMSDs below 10 Å, identified their corresponding MSAs, and compiled a focused MSA subset enriched in sequences capable of predicting the inactive state. This enriched MSA, comprising 2,979 sequences, was then used for M-fold sampling.

In the first round of M-fold sampling, structural predictions remained biased toward the active state, with only a small fraction exhibiting low RMSD relative to the inactive state (**Fig. 4b**). The RMSD range (0-25 Å) was divided into 30 bins, and sequence voting identified the 4$^{th}$ bin as the purified MSA subsets for the active state (RMSD relative to the DFG-in conformation: ~3.0 Å), but the low representation of DFG-out structures complicated the voting process. To overcome this limitation, we recollected sequences producing structures with RMSDs below 10 Å and conducted a second round of M-fold sampling. This round utilized a reduced MSA (110 sequences) with 100 rounds of reshuffling and sampling, substantially increasing the number of predicted structures corresponding to the inactive state (**Fig. 4c**). The structural distribution after this refinement shifted markedly toward the inactive state compared to the first round (**Fig. 4d**).

Subsequent sequence voting identified bins 7, 8, and 9 as purified MSA subsets for the inactive state (**Fig. 4c**). Using these purified MSAs, we successfully predicted the inactive conformation with an A-Loop RMSD of 4.6 Å relative to the DFG-out state (**Fig. 4e**). Remarkably, even for bin 7 with only 13 purified sequences, all of predicted structures converged to the same inactive state with small variation (RMSD: 4.6-5.8 Å). In contrast, predictions using randomly selected sequences from the raw MSA consistently failed to capture the inactive state and only produced the active conformation. These findings demonstrate that structural prediction variability depends on the purity of the evolutionary signal encoded in the MSA rather than sequence depth.

Similar to AdK, purified sequences capable of predicting the inactive state were sparsely distributed across evolutionary space. As shown in **Extended Data Fig. 8**, sequences in bins 7, 8, and 9 did not cluster within a single clade or branch of the phylogenetic tree. Furthermore, their sequence identity to the query sequence varied widely, indicating that structural prediction accuracy is not strictly dependent on close evolutionary proximity or sequence similarity.

**Predicting active and inactive states of GLP-1R**

The glucagon-like peptide-1 receptor (GLP-1R), a G-protein-coupled receptor (GPCR), undergoes conformational transitions between active and inactive states in response to glucagon-like peptides[55-58]. Activation is marked by an increase in the distance between the cytosolic ends of transmembrane helices TM3 and TM6, which expands to approximately 37 Å as TM3 tilts, enabling G-protein binding. In the inactive state, this distance contracts to ~18 Å[55, 59, 60] (**Fig. 5a**). This TM3-TM6 distance serves as a reliable feature for distinguishing between states during M-fold sampling.

Default AF2 predictions using raw MSAs failed to capture these well-defined states, instead producing a mixture of conformations spanning the active and inactive states (**Fig. 5b**). This suggests that noisy evolutionary signals in the raw MSA obscure critical structural distinctions. Applying M-fold sampling to GLP-1R revealed a continuous distribution of distinct conformational states, with TM3-TM6 distances ranging from 15 to 40 Å (**Fig. 5c**). By voting sequences based on TM3-TM6 distances, we identified specific bins associated with each conformational state - bins 30 and 31 for the inactive state and bin 77 for the active state.

Predicted structures using purified sequences showed remarkable accuracy, reliably converging on their respective TM3-TM6 distances of 33 Å (transmembrane domain TM-score compared to the active state as 0.94) and 21 Å (transmembrane domain TM-score compared to the inactive state as 0.89) (**Figs. 5d, 5e**). In contrast, predictions using subsampled unpurified MSAs exhibited a broader distribution of conformations, with TM3-TM6 distances ranging from 20 to 33 Å. These findings align with previous studies suggesting that reducing the depth of MSAs can promote the generation of alternative conformations[20, 24, 29]. However, our results reveal that this increased diversity does not necessarily yield accurate structures for most proteins. As shown in **Fig. 5c**, the conformational distribution of GLP1R is relatively uniform and continuous across the entire TM3-TM6 distance space, allowing the sampling of diverse conformations, including both active and inactive states. By contrast, for proteins like AdK or ABL1, reducing MSA depth fails to produce alternative structures, highlighting the limitations of this strategy for sampling rare conformational states.

Interestingly, although the extracellular domain (ECD) was not explicitly considered during M-fold sampling, purified sequences associated with the inactive state often predicted diverse ECD conformations, reflecting the domain's structural plasticity in the absence of GLP peptides (**Fig. 5d**). In contrast, purified sequences corresponding to the active state produced highly convergent ECD structures, potentially critical for peptide binding. To explore the evolutionary origins of these purified sequences, we mapped them onto the phylogenetic tree of GLP-1R sequences. Consistent with observations in other systems, these sequences were not restricted to close evolutionary homologs of the query (**Extended Data Fig. 9**).

**Predicting fold-switch and ground states of KaiB with purified sequences**

KaiB, a pivotal component of the cyanobacterial circadian clock, exhibits dynamic structural transitions between a fold-switch state and a ground state, integral to its role in regulating rhythmic oscillations[61-63] (**Fig. 6a**). These conformational shifts enable KaiB to interact with other clock proteins, orchestrating the periodicity of the circadian system. Accurately predicting KaiB's distinct structural states, particularly its ground state, remains a significant challenge for AF2[28], given the sparse representation of ground-state-associated sequences in evolutionary data[28].

Similar to the challenges faced in predicting rare inactive states of ABL1, identifying sequences contributing to KaiB's ground state required iterative shuffling and enrichment prior to M-fold sampling (**Extended Data Fig. 7b**). During each iteration, MSA subsets that yielded predicted structures with TM-scores exceeding the mean by +0.03 were selected for further refinement. After eight iterations, we collected all structures with TM-scores above 0.65, and identified their corresponding MSAs, which were compiled into a focused MSA subset for subsequent M-fold sampling. This iterative enrichment revealed two distinct high-density structural clusters corresponding to the ground state and fold-switch state (**Fig. 6b**). Sequence voting on the enriched MSA further resolved these states, with bin 20 (**Fig. 6c**, left panel) containing sequences associated with the ground state, while bin 26 (**Fig. 6c**, right panel) corresponded to the fold-switch state.

Predictions using purified sequences consistently converged onto a single state, achieving TM-score of 0.8 for the fold-switch state and 0.92 for the ground state (**Figs. 6d, 6e**). By contrast, predictions using randomly selected sequences from the raw MSA exhibited high variability and failed to capture the ground state entirely. Various methods, including AF_Cluster and other[18, 28, 64], have successfully predicted alternative KaiB structures. However, our approach offers a distinct advantage by not only resolving both the ground and fold-switch states with high confidence but also pinpointing the specific sequences contributing to each state (**Extended Data Fig. 10**).

Intriguingly, during M-fold sampling, a distinct subset of MSAs emerged, corresponding to conformations that did not align with either the fold-switch or ground states. Filtering structures with pLDDT scores above 85 and applying principal component analysis (PCA) to Cα atoms revealed an extended conformation distinct from the canonical states (**Fig. 6f**). Previous NMR studies have suggested a conformational transition pathway for KaiB, in which β3' and α4' undergo partial unfolding before switching from the fold-switch state to the ground state.[65] The newly identified conformation closely resembles an intermediate along this pathway, retaining structural elements preceding β2 while exhibiting partial unfolding of β3' and α4'. This suggests the existence of a previously unidentified intermediate bridging the fold-switch and ground states, expanding the known conformational ensemble of KaiB. These findings demonstrate that AF-ClaSeq can extend beyond capturing well-defined conformations to uncover novel intermediate states, offering deeper insights into protein structural dynamics and functional mechanisms.

### Predicting 3α and 4β1α conformations of *de novo* designed proteins $G_A98$ and $G_B98$

To evaluate its performance on *de novo* designed proteins, we applied AF-ClaSeq to two synthetic mini-proteins, $G_A98$ and $G_B98$, which differ by a single residue at position 45 (L45 in $G_A98$ and Y45 in $G_B98$), yet adopt entirely different structural folds[66]. This single substitution triggers a conformational shift from a 3α helix to a 4β1α fold (**Fig. 7a**). AF2 predictions for $G_B98$ using default MSAs were inconsistent. Of the five predicted structures, only two adopted the correct fold (Cα RMSD relative to the experimental structure: 1.59 Å), while the other three exhibited incorrect structures with Cα RMSD values exceeding 9 Å (**Fig. 7b**). To refine these predictions, we applied M-fold sampling on the raw MSA (**Fig. 7c**). Although the source MSA was derived from the $G_B98$ sequence, initial M-fold sampling yielded far more 3α-like structures than the expected 4β1α fold. To offset this imbalance, we selected a subset of structures with an RMSD below 2.5 Å relative to the 4β1α conformation and performed a second round of M-fold sampling. Subsequent sequence voting identified bins 4 and 5 (left panel in **Fig. 7d**) as corresponding to the 3α state and bin 5 (right panel in **Fig. 7d**) to the 4β1α state. Using these purified sequences, AF2 accurately predicted both conformations with high confidence, achieving RMSD values below 2 Å for the α conformation and as low as 1.8 Å for the β conformation.

Beyond the L45/Y45 substitution, previous experiments have implicated additional mutations in modulating fold switching. Specifically, the $G_B98$-T25I mutation induces the 3α conformation, while the combined $G_B98$-T25I/L20A mutations restore the 4β1α fold. We generated 80 structures per mutant using AF2 with 16 different random seeds. Default MSAs often failed to yield the correct structures (**Extended Data Fig. 11**). However, when, predictions using purified fold-specific sequences consistently converged on the correct structures, with RMSD values below 2.5 Å relative to experimental structures.

### Insights from purified MSAs

We next examine whether purified MSAs can uncover key insights into a protein's structural dynamics and function, which are often obscured by noise in unpurified MSAs.

For adenylate kinase (AdK), AF-ClaSeq identified two distinct MSA subsets corresponding to the apo and ligand-bound states. To characterize the differences between these subsets, we transformed them into residue pairwise representations. Specifically, we extracted embeddings from the Evoformer output - without iterative recycling[1, 44, 47]. Each MSA is represented as a 3D tensor of dimensions ($N_{_res}$, $N_{_res}$, c), where $N_{_res}$ denotes the number of residues and c = 128 represents residue pair embeddings derived from the MSA (**Fig. 8a**). A difference tensor was then computed to capture variations between the two MSAs. To extract contrastive features between the apo and ligand-bound MSAs, we applied PCA and uniform manifold approximation and projection (UMAP) to reduce the dimensionality of the pairwise representations. The reduced 2D map visualizes each residue pair as a point, color-coded by its distance difference between the two states (**Fig. 8b**), allowing for mapping MSA-derived features onto structural transitions. Residue pairs with PC1 values exceeding 150 (shown as yellow dashed lines) were mapped onto the apo AdK structure (**Fig. 8c**), revealing key residues implicated in the conformational change. Notably, residue pairs exhibiting substantial differences in inter-residue distances between states - indicative of dynamic relevance - were effectively distinguished using either PCA or UMAP. These findings provide a mechanistic explanation for why purified MSAs consistently predict distinct conformational states as each MSA subset contains residue pair information specific to a given state. Importantly, as these representations were obtained from Evoformer embeddings without any recycling[1, 44, 47], the insights arise entirely from sequence data, independent of any structural input.

We extended our analysis to the $G_A98$/$G_B98$ proteins to identify key residue correlations governing their conformational preferences. We selected three representative, conformation-specific residue pairs for each state from a reduced 2D map that revealed residue pairs involved in the fold switch between the 3α and 4β1α conformations. These residue pairs were marked with red asterisks on their respective contact maps (**Fig. 8d**). To further investigate their sequence-level contributions, we generated residue occupancy heatmaps from the corresponding purified MSAs, illustrating the frequency of specific amino acids at each position for a given residue pair. Mapping these residues onto the $G_A98$/$G_B98$ structures revealed that $G_A98$-exclusive residue pairs exhibited strong complementarity within the $G_A98$-purified MSA but appeared conflicting or noisy in the $G_B98$-purified MSA, and *vice versa* for $G_B98$-exclusive residue pairs (**Fig. 8d**). This analysis underscores the effectiveness of sequence purification in extracting co-evolutionary signals embedded in MSAs, offering unique insights into protein structural dynamics from a sequence-based perspective.

## Discussion

We present AF-ClaSeq, a sequence purification method that enhances AlphaFold2 (AF2)'s ability to predict multiple conformational states with improved accuracy. AF-ClaSeq has demonstrated effectiveness across a diverse range of proteins, capturing structural variations that encompass global fold-switching, domain motions, and local loop dynamics. Purified MSA subsets offer more than just improved structural predictions - they provide insights into critical "hotspot" residues and co-evolutionary couplings that govern specific conformational states and transitions within protein families.

Existing methods for predicting alternative structures rely on MSA depth reduction, clustering, or masking[21-23, 25, 28, 64, 67]. Our findings reveal that purified MSA subsets are not confined to specific clusters but are instead dispersed across phylogenetic trees. This observation suggests that co-evolutionary signals dictating conformational preferences cannot be captured by clustering or MSA depth reduction. Rather, sequence classification offers a viable solution by identifying subsets of sequences exhibiting distinct structural preferences. However, exhaustive searches through all possible MSA subsets are computationally prohibitive, and the lack of co-evolutionary depth in individual sequences further complicates classification. To address this, AF-ClaSeq employs a structural-guided M-fold sampling and voting strategy, akin to AlphaCryo4D[38], to enhance signal-to-noise ratio and extract meaningful co-evolutionary information.

Recent work suggested that co-evolutionary information has limited capability for sampling alternative conformations[26]. In contrast, our results demonstrate that AF2, when provided with purified MSA subsets, even with a small number of sequences, can produce high-confidence predictions with low structural variability. The apparent disconnect between AF's internal embeddings and structural outputs likely arises from noisy MSAs. By systematically isolating sequences encoding distinct conformational states, AF-ClaSeq overcomes this limitation, enabling AF2 to resolve alternative structures with enhanced accuracy and confidence.

Most existing AF2-based methods for predicting alternative structures rely on prior knowledge of these structures[26, 27], constrained by the lack of reliable criteria for identifying unknown structures. Metrics like pLDDT scores, while indicative of prediction confidence, are insufficient for distinguishing native structures from decoys[68, 69]. However, we found that pLDDT scores correlate strongly with the purity of MSAs, allowing for high-confidence predictions of novel structures. While some cases, such as ABL1 and KaiB, utilized the coordinates of both conformational states to guide sequence purification, this information is not strictly necessary. Loosely defined reaction coordinates - such as inter-domain distances, pseudo-hinge angles, or latent-space features derived from neural networks - can effectively direct sequence purification. This flexibility allows AF-ClaSeq to not only recapitulate known structures but also predict novel conformations that could even be transient or metastable.

While generally effective, AF-ClaSeq exhibits variability in performance across different systems. For example, it successfully predicted active, inactive, and intermediate states of GLP1R, but initially failed to resolve the inactive conformation of ABL1 kinase due to sparsely populated MSA subsets encoding this rare state. The skewed distribution of these sequences, often associated with ancient or functionally obsolete proteins, hampers the extraction of co-evolutionary information. To overcome this, we implemented an iterative enrichment strategy to bias sequence selection toward the rare conformational state, ultimately enabling accurate predictions of multiple Abel kinase conformations, including subtle structural rearrangements in the DFG loop.

Proteins execute their functions through conformational changes triggered by cellular events, such as ligand binding, protein-protein interactions, or post-translational modifications. While protein structures tend to be evolutionarily conserved, their conformational dynamics often diverge to meet distinct functional requirements. Consequently, co-evolutionary patterns encoding these functionally relevant dynamics are often obscured in raw MSAs. By leveraging structure-guided purification, AF-ClaSeq isolates these signals, enabling the prediction of protein dynamics with greater accuracy. Notably, our findings suggest that protein dynamics can be predicted independently of functional context, provided that the underlying co-evolutionary patterns are extracted through purification. The resulting purified MSA subsets open opportunities for predicting alternative conformations that a protein may not normally adopt but could assume under specific functional conditions, such as in the presence of a binding partner, post-translational modification, or within a particular biological environment. This capability underscores the potential of AF-ClaSeq to uncover hidden structural plasticity, broadening our understanding of protein dynamics and expanding the predictive power of AF2-based structural modeling.

**Acknowledgements**

This work was supported by the funding from the National Institutes of Health (NIH) under the Award No R01GM141394 and from the National Science Foundation (NSF) under the Award No CHE-2402592.

**Author Contributions**

X.C. and E.X. conceptualized the work. E.X., J.Z. and S.W. performed research and analyzed data. E.X. and X.C. led the writing of the manuscript with contributions from all co-authors.

## Methods

### Multiple Sequence Alignment (MSA) and structure prediction

The multiple sequence alignments (MSAs) for all protein cases used in this study were obtained from either the MMseqs2 server[42] implemented in ColabFold v1.5.5[44, 47] or the DeepMSA2[45, 46] package, which performed sequence searches against databases including UniClust30[70], UniRef90[71], MetaClust[39], MGnify[72], BFD[39], TaraDB[46], MetaSourceDB[46, 73], and JGIclust[46, 74]. To ensure quality and mitigate noise from gap regions, the source MSA may be trimmed based on sequence coverage. All structure predictions, whether from source MSA or their subsets, were carried out using the ColabFold implementation of AlphaFold2. During M-fold sampling, only one structure was generated for each MSA subset file, with the random seed set to 42. For each purified MSA subset, 8 random seeds were given, and 5 models were generated per random seed, yielding a total of 40 predicted structures to assess the accuracy and variability of the predictions. The number of recycles were fixed at 3 across all predictions. The commands used for structure prediction are listed below:

- For predictions during M-fold sampling:

```
colabfold_batch input_dir_with_msas/ output_dir/ --num-models 1 --random-seed 42 --num-recycle 3
```

- For predictions after sequence purification:

```
colabfold_batch input_dir_with_msas/ output_dir/ --num-models 5 --num-seeds 8 --num-recycle 3
```

### Structural metrics calculation

Key structural metrics used for the classification of protein conformational states included TM-score, Cα atom Root Mean Square Deviation (RMSD), and domain distance. TM-Score was calculated using the TM-Align executable obtained from Dr. Yang Zhang's lab[75]. For RMSD calculation, protein Cα atom coordinates were extracted and superimposed using the *superimposer* module from the Biopython[76] software package. Residue indices within the domain of interest were identified, and the corresponding Cα atom coordinates were used to compute RMSD based on the following equation:

$$RMSD = \sqrt{\frac{1}{N}\sum_{i=1}^{N}\left\|\mathbf{r}_i^{ref} - \mathbf{r}_i^{target}\right\|^2}$$

where $N$ is the number of Cα atoms considered, and $\mathbf{r}_i^{ref}$ and $\mathbf{r}_i^{target}$ denote the Cartesian coordinates of the $i$-th Cα atom in the reference and target structures, respectively.

Inter-domain distance was determined by first calculating the centers-of-mass for two distinct sets of residues within a specified domain. For a group of $M$ atoms, the center-of-mass $\mathbf{c}$ is given by

$$\mathbf{c} = \frac{\sum_{i=1}^{M} m_i \mathbf{r}_i}{\sum_{i=1}^{M} m_i}$$

where $m_i$ and $\mathbf{r}_i$ are the mass and coordinate vector of the $i$-th atom. The Euclidean distance $d$ between the centers-of-mass of the two residue groups is then calculated as:

$$d = \|\mathbf{c}_1 - \mathbf{c}_2\|$$

**Iterative shuffling and enrichment of rare sequences in MSA**

The iterative sequence enrichment process begins with a raw multiple sequence alignment (MSA) as input. In each iteration, sequences are randomly shuffled and split into fixed-size groups, with the process repeated ten times to ensure sampling diversity. Each resulting MSA subset is the subjected to structure prediction using ColabFold, followed by quantitative structural analysis based on metrics such as RMSD to reference conformations or specific inter-residue distances. Structures that meet predefined criteria (e.g., RMSD below the 20th percentile) are identified, and their corresponding sequences are collected from the source MSAs. These sequences are recompiled into a new MSA subset, with duplicates removed. The filtered sequence pool is then subject to subsequent iterations of shuffling, prediction, and selection, progressively enriching for sequences that favor specific conformational states. After multiple iterations - determined on a case-by-case basis - sequences corresponding to all qualifying structures are compiled into a final dataset for M-fold sampling.

**M-Fold sampling and sequence voting**

The filtered MSA is partitioned into M+1 groups, each containing N sequences. A leave-one-group-out bootstrapping procedure is applied, systematically excluding one group while redistributing the remaining sequences into M new groups, each retaining N sequences. This procedure generates M×(M+1) distinct MSA subsets, ensuring that each sequence appears exactly M times across all subsets.

Each MSA subset is subsequently subjected to structure prediction using ColabFold, producing M×(M+1) predicted structures. These structures are sorted into bins based on a predefined reaction coordinate (e.g., root-mean-square deviation (RMSD) to reference structures or other collective variables). This process yields a probability distribution reflecting the conformational landscape in the sequence space. By partitioning the initial MSA into smaller subsets, the approach amplifies the contributions of individual sequences to specific conformational states, revealing subtle sequence-structure relationships.

To identify sequences that preferentially contribute to distinct structural states, a voting mechanism was implemented. For each sequence, its presence across M groups and the corresponding predicted structures is tracked. Predicted structures are binned according to the chosen reaction coordinate, and voting statistics are calculated by analyzing the frequency of bin assignments. A sequence is ultimately assigned to the bin where its associated structures appear most frequently. This statistical approach enhances the contributions of individual sequences to ensemble structural predictions, effectively identifying structure-specific sequences from a large and noisy sequence pool. Finally, sequences linked to specific structural states are compiled into new MSAs, referred to as purified MSAs for those states.

## Figure Legend

**Fig. 1. Sequence purification enhances AlphaFold2's ability of predicting multiple conformational states**. **a**. A raw MSA, containing sequences with noisy or conflicting co-evolutionary information, may fail to capture residue-residue couplings necessary for accurate structural prediction. This often leads to biased predictions favoring a dominant conformational state (e.g., domain motion) or produces ensemble-averaged structures (e.g., loop motion) with low confidence; **b**. Sequence purification addresses this limitation by grouping sequences that bias toward the same specific state into a refined MSA subset. This process resolves frustrations hidden in the raw MSA, enabling AlphaFold2 to predict distinct conformational states with high confidence.

**Fig. 2. Overview of the AF-ClaSeq pipeline for structure-based sequence purification. a**. The pipeline begins with a query sequence processed through DeepMSA2 or MMSeqs2 to generate a comprehensive multiple sequence alignment (MSA). Depending on the size and distribution of the MSA, a filtering or enrichment step may be applied to prune the MSA; **b**. The initial or trimmed MSA undergoes M-fold sampling to generate M×(M+1) distinct MSA subsets, each containing N sequences. These subsets are individually input into AlphaFold2 to predict structures. Predicted structures are grouped into bins based on predefined structural metrics (or reaction coordinates). A voting mechanism is then applied to calculate the probability of each sequence contributing to the prediction of distinct conformational states. Sequences that bias toward the same specific state are then grouped into a refined, or purified, MSA subset, enabling AlphaFold2 to predict the corresponding conformational state with high confidence.

**Fig. 3. Sequence purification for adenylate kinase. a**. Experimentally determined structures of the apo state (green) and ligand-bound state (purple), alongside AlphaFold2-predicted structures (grey) using the default MSA. Predicted structure are superimposed on both experimental states for comparison; **b**. RMSD distribution of all predicted structures during M-fold sampling, color-coded by pLDDT scores. **c**. Probability distributions of RMSD values for predicted structures relative to the apo and bound states (top row). Voting probability distributions of sequences assigned to distinct RMSD bins relative to each state (bottom row). **d.** 2D RMSD plot comparing structures predicted using purified sequences for each state versus those predicted using randomly selected sequences, color-coded by pLDDT scores; **e.** Superimposition of the top 10 structures ranked by pLDDT scores for the apo (left) and ligand-bound (middle) states, with the highest-ranked structure aligned to the corresponding experimental structure. Structures predicted using randomly selected sequences are shown for comparison (right). Proteins are represented as ribbons and color-coded by pLDDT scores.

**Fig. 4. Sequence purification for ABL1 kinase. a**. Experimentally determined structures of the active state and inactive state, alongside AlphaFold2-predicted structure using the default MSA (A-Loop highlighted in green, purple, and yellow, respectively). The predicted structure with the highest pLDDT score is superimposed on both experimental structures, revealing that the default prediction captures only the active DFG-in state but fails to predict the inactive DFG-out state; **b**. Probability distributions of A-Loop RMSD values for predicted structures relative to the active and inactive states during the first round of M-fold sampling (top row). Voting probability distributions of sequences assigned to distinct RMSD bins relative to the active and inactive states (bottom row); **c**. Second round of M-fold sampling for sequences that predicted inactive structures (RMSD < 10 Å) in the first round; **d**. 2D RMSD plots of predicted structures from the first and second rounds of M-fold sampling, color-coded by global pLDDT scores; **e**. 2D RMSD plots comparing structures predicted from purified sequences versus those generated using randomly selected sequences, color-coded by global pLDDT scores. The top 10 structures with the highest pLDDT scores are superimposed (inset), with proteins shown as ribbons and color-coded by pLDDT scores.

**Fig. 5. Sequence purification for GLP1R. a**. Experimentally determined structures of the active (purple) and inactive (blue) states. The TM3-TM6 distance is defined as the center distance between the cytosol termini of the TM3 and TM6 helices. **b**. Comparison of default AlphaFold2 predicted structures with both experimental states. Predictions using the raw MSA fail to resolve either state clearly, instead exhibiting significant variability in the TM6 helix; **c.** Probability distributions of TM3-TM6 distances for predicted structures during M-fold sampling (top row). Voting probability distributions for sequences assigned to distinct TM3-TM6 distance bins (bottom row). **d**. 2D TM-score/TM3-TM6 distance plots comparing structures predicted from purified sequences versus those generated using randomly selected sequences, color-coded by global pLDDT scores. The top 10 structures with the highest pLDDT scores are superimposed (inset), with proteins shown as ribbons and color-coded by pLDDT scores. **e**. Predictions from purified MSAs corresponding to the active and inactive states show clear convergence toward a single state.

**Fig. 6. Sequence purification for KaiB. a**. Experimentally determined structures of the fold-switch state (green) and ground state (purple), alongside AlphaFold2-predicted structures (grey) using the default MSA. Default predictions capture only the fold-switch state. **b.** 2D map of predicted structures based on TM-scores relative to the fold-switch and ground states during M-fold sampling; **c**. TM-score probability distributions for predicted structures relative to the fold-switch and ground states during M-fold sampling (top row). Voting probability distributions for sequences assigned to distinct TM-score bins corresponding to each state (bottom row); **d.** 2D maps of structures predicted from purified MSAs corresponding to the fold-switch and ground states, compared to those generated using randomly selected sequences; **e.** Superimposition of the top 10 structures ranked by pLDDT scores for the fold-switch state (left) and ground state (middle), with the highest-ranked structure aligned to the corresponding experimental structure. Structures predicted using randomly selected sequences are shown for comparison (right). Proteins are represented as ribbons and color-coded by pLDDT scores. **f.** 2D TM-score map of predicted structures during M-fold sampling reveals a significant population of a previously uncharacterized state with low TM-scores relative to both the fold-switch (<0.65) and ground (<0.52) states but a high pLDDT score (> 80). Ten representative structures are superimposed (left), with the highest-ranked structure aligned to the fold-switch (middle) and ground (right) states. This structure aligns well with the fold-switch state except for the partial unfolding in β3' and α4'.

**Fig. 7. Sequence purification for synthetic mini-proteins $G_A$98 and $G_B$98. a**. $G_A$98 and $G_B$98 differ by a single amino acid at position 45. Experimentally determined structures of $G_A$98 (green) and $G_B$98 (purple) are shown with secondary structure labelled. **b**. AlphaFold2-predicted structures (grey) using the default MSA of $G_B$98 sequence. 5 models were generated and superimposed onto the experimental $G_B$98 structure for comparison. **c.** RMSD probability distributions of predicted structures relative to the $G_A$98 (3α) and $G_B$98 (4β1α) conformations during the first round of M-fold sampling. Sequences contributing to structures within the pink-shaded region were selected and compiled for the second round of M-fold sampling. **d.** RMSD probability distributions of predicted structures during the second round of M-fold sampling (top row). Voting probability distributions for sequences assigned to distinct RMSD bins corresponding to the 3α and 4β1α states (bottom row); **e.** 2D RMSD maps of predicted structures during the first and second rounds of M-fold sampling, color-coded by pLDDT scores; **f.** 2D TM-score plots comparing structures predicted from purified sequences versus those generated using randomly selected sequences, color-coded by global pLDDT scores. The top 10 structures with the highest pLDDT scores are superimposed (inset), with proteins shown as ribbons and color-coded by pLDDT scores.

**Fig. 8. Insights from purified sequences. a.** Pairwise representations of adenylate kinase for purified MSA subsets corresponding to the apo and ligand-bound states. The 3D tensors have dimensions of *($N_{res}$, $N_{res}$, c),* where $N_{res}$ represents the number of residues, and *c* = 128 corresponds to the Evoformer embeddings for individual residue pairs derived from AlphaFold2. A difference tensor highlights variations between the two states. Residue pair embeddings were subjected to dimensionality reduction using PCA and UMAP, generating reduced 2D maps in which each residue pair is represented as a point, color-coded by inter-residue distance differences. **b.** Residue pairs with PC1 values >150 are highlighted with yellow dashed lines in the structural representation. **c.** Residue contact maps for $G_A98$ (top) and $G_B98$ (bottom), with three representative state-specific residue pairs highlighted by red asterisks. Amino acid heatmaps derived from purified MSAs for $G_A98$ (left) and $G_B98$ (right) are shown for the three $G_A98$-specific (top) and three $G_B98$-specific (bottom) residue pairs. These pairs are also highlighted in the $G_A98$ and $G_B98$ structural models.

**Extended Data Fig. 1. Comparison of structural prediction performance between AF-ClaSeq and AF_Cluster across eight protein systems. a.** RMSD or TM-score distributions of predicted structures using MSA subsets derived from AF-ClaSeq (left column) and AF_Cluster (middle and right columns). The middle column displays the full distribution of all predicted structures from AF_Cluster-derived MSAs, while the right column provides a zoomed-in view scaled to match the distribution from AF-ClaSeq for direct comparison. **b.** RMSD or TM-score distributions of predictions for specific conformational states using MSA subsets identified by AF_Cluster. Structures most resembling the two states of each protein (determined by the lowest RMSD or highest TM-Score) were identified, and their corresponding MSAs were retrieved. AlphaFold2 predictions were then performed using these MSAs, generated 40 structures with 8 random seeds. These resulting structures were mapped onto RMSD or TM-score 2D plots to visualize predictive accuracy.

**Extended Data Fig. 2. Sequence purification for AtaR toxin. a**. Experimentally determined structures of the neutralizing state (green) and non-neutralizing state (purple) of AtaR toxin, alongside AlphaFold2-predicted structures (grey) using the default MSA. The experimental structure of the neutralizing state covers residue LYS6 to GLN86, while the non-neutralizing state covers residue ILE9 to ASN70; **b**. 2D map of predicted structures based on RMSD values relative to the neutralizing and non-neutralizing states during M-fold sampling; **c**. RMSD probability distributions for predicted structures relative to the neutralizing (left) and non-neutralizing (right) states (top row). Voting probability distributions for sequences assigned to distinct bins corresponding to each state (bottom row). **d.** 2D maps of structures predicted from purified MSAs corresponding to the neutralizing and non-neutralizing conformations, compared to those generated using randomly selected sequences; **e.** Superimposition of the top 10 structures ranked by pLDDT scores for the neutralizing state (left) and non-neutralizing state (middle), with the highest-ranked structure aligned to the corresponding experimental structure. Structures predicted using randomly selected sequences are shown for comparison (right). Proteins are represented as ribbons and color-coded by pLDDT scores.

**Extended Data Fig. 3. Sequence purification for human H-Ras protein. a**. Experimentally determined structures of the GMPNP-bound active (green) and GDP-bound inactive (purple) states of human H-Ras protein, alongside AlphaFold2-predicted structures (grey) using the default MSA. Switch-I loop is highlighted in superimposed structures; **b**. Default AlphaFold2-predicted structures of H-Ras are shown as ribbons and color-coded by local pLDDT scores. The Switch-I loop exhibits low pLDDT scores, reflecting its high conformational plasticity and dynamic characteristics. **c**. Enrichment of sequences favoring the GDP-bound inactive state. The red dashed lines indicate the RMSD threshold for the lowest 20% Switch-I conformations relative to the inactive state. Structures within the pink-shaded region were selected, and their corresponding MSAs were identified. Unique sequences in these MSAs were compiled for M-fold sampling; **d.** 2D RMSD maps of predicted structures relative to the active and inactive states during M-fold sampling, color-coded by either global (left) and local Switch-I loop (right) pLDDT scores. **e**. RMSD probability distributions for predicted structures relative to the active switch-I (left) and inactive switch-I (right) states (top row). Voting probability distributions for sequences assigned to distinct bins corresponding to each state (bottom row). **f.** 2D RMSD maps of structures predicted from purified MSAs corresponding to the active switch-I and inactive switch-I states, compared to those generated using randomly selected sequences; **g.** Superimposition of the top 10 structures ranked by pLDDT scores for the active switch-I (left) and inactive switch-I (middle) states, with the highest-ranked structure aligned to the corresponding experimental structure. Structures predicted using randomly selected sequences are shown for comparison (right). Proteins are represented as ribbons and color-coded by pLDDT scores.

**Extended Data Fig. 4. Sequence purification for family II inorganic pyrophosphatase. a**. Experimentally determined structures of the closed (green) and open (purple) state family II inorganic pyrophosphatase protein, alongside AlphaFold2-predicted structures (grey) using the default MSA; **b**. 2D map of predicted structures based on TM-scores relative to the closed and open states during the first round of M-fold sampling. Structures with TM-scores relative to the closed state > 0.86 and TM-scores relative to the open state > 0.82 were collected, highlighted by green and pink shadowed region. MSAs corresponding to these structures were compiled and subjected to focused M-fold sampling; **c**. TM-score probability distributions of predicted structures relative to the closed and open conformations during the first round of M-fold sampling. Sequences contributing to structures within the green- and pink-shaded regions were selected and compiled for the second round of M-fold sampling. **d**. TM-score probability distributions of predicted structures during the focused M-fold sampling (top row). Voting probability distributions for sequences assigned to distinct TM-score bins corresponding to the closed and open states (bottom row). **e.** 2D map of predicted structures based on TM-scores relative to the closed and open states during the focused M-fold sampling on each state (top row). 2D TM-score maps of structures predicted from purified MSAs corresponding to the closed and open states (middle row), compared to those generated using randomly selected sequences (bottom row); **f.** Superimposition of the top 10 structures ranked by pLDDT scores for the closed (left) and open (middle) states, with the highest-ranked structure aligned to the corresponding experimental structure. Structures predicted using randomly selected sequences are shown for comparison (right). Proteins are represented as ribbons and color-coded by pLDDT scores.

**Extended Data Fig. 5**. Sequence coverage analysis and visualization of the evolutionary distance among the purified sequences of adenylate kinase. The purified sequences in the $9^{th}$ bin (shown as blue lines in the phylogenetic trees), corresponding to the apo state, and those in the $8^{th}$ bin (shown as purple lines in the phylogenetic trees), corresponding to the ligand-bound state, are not closely clustered and do not belong to the same clade.

**Extended Data Fig. 6. a**. Dimensionality reduction of Cα atom coordinates for adenylate kinase using a ResNet-based variational autoencoder (VAE). All predicted structures obtained during M-fold sampling are projected onto a UMAP in the reduced 2D feature space. **b**. 2D voting probability distributions for sequences assigned to distinct bins based on two UMAP coordinates; **c**. Predicted structures from purified MSA subsets corresponding to distinct intermediate states between apo and ligand-bound conformations; **d.** Superimposition of predicted structures representing distinct states. **e.** Prediction results from all UMAP bins containing more than 20 sequences.

**Extended Data Fig. 7**. **a**. Enrichment of sequences favoring the inactive state. The red dashed lines indicate the RMSD threshold for the lowest 20% A-loop conformations relative to the inactive state. Structures within the pink-shaded region were selected, and their corresponding MSAs were identified, of which unique sequences were compiled for M-fold sampling; **b**. Iterative enrichment of sequences favoring the ground state. The red dashed lines indicate the TM-score threshold (>0.65) for structures closely matching the ground state. Structures within the pink-shaded region were selected, and their corresponding MSAs were identified. Unique sequences in these MSAs were compiled for M-fold sampling.

**Extended Data Fig. 8**. Sequence coverage analysis and visualization of the evolutionary distances among the purified sequences of ABL1 kinase. The purified sequences in the $7^{th}$, $8^{th}$, and $9^{th}$ bins (shown as blue lines in the phylogenetic trees), corresponding to the inactive state, are not closely clustered and do not belong to the same clade.

**Extended Data Fig. 9**. Sequence coverage analysis and visualization of the evolutionary distance among the purified sequences of GLP1R. The purified sequences in the $30^{th}$ and $31^{st}$ bins (shown as blue lines in the phylogenetic trees), corresponding to the inactive state, and those in the $77^{th}$ bin (shown as purple lines in the phylogenetic trees), corresponding to the active state, are not closely clustered and do not belong to the same clade.

**Extended Data Fig. 10**. Sequence coverage analysis and visualization of the evolutionary distance among the purified sequences of KaiB. The purified sequences in the $26^{th}$ bin (shown as blue lines in the phylogenetic trees), corresponding to the ground state, and those in the $20^{th}$ bin (shown as purple lines in the phylogenetic trees), corresponding to the fold-switch state, are not closely clustered and do not belong to the same clade.

**Extended Data Fig. 11.** Comparison of AlphaFold2 prediction results for $G_A98$, $G_B98$, and two $G_B98$ mutants using either the default MSA queried from the MMseqs2 server (top) or the purified MSAs corresponding to the 3α and 4β1α states. Sixteen random seeds were used, with each generating five models, resulting in a total of 80 structures per case. The Cα atom RMSDs of all predicted structures were computed against their respective experimental structures and visualized by a strip plot. The purified MSA subsets for the 3α and 4β1α states are shown in the right panel, highlighted in green and purple, respectively.

# Figure 1

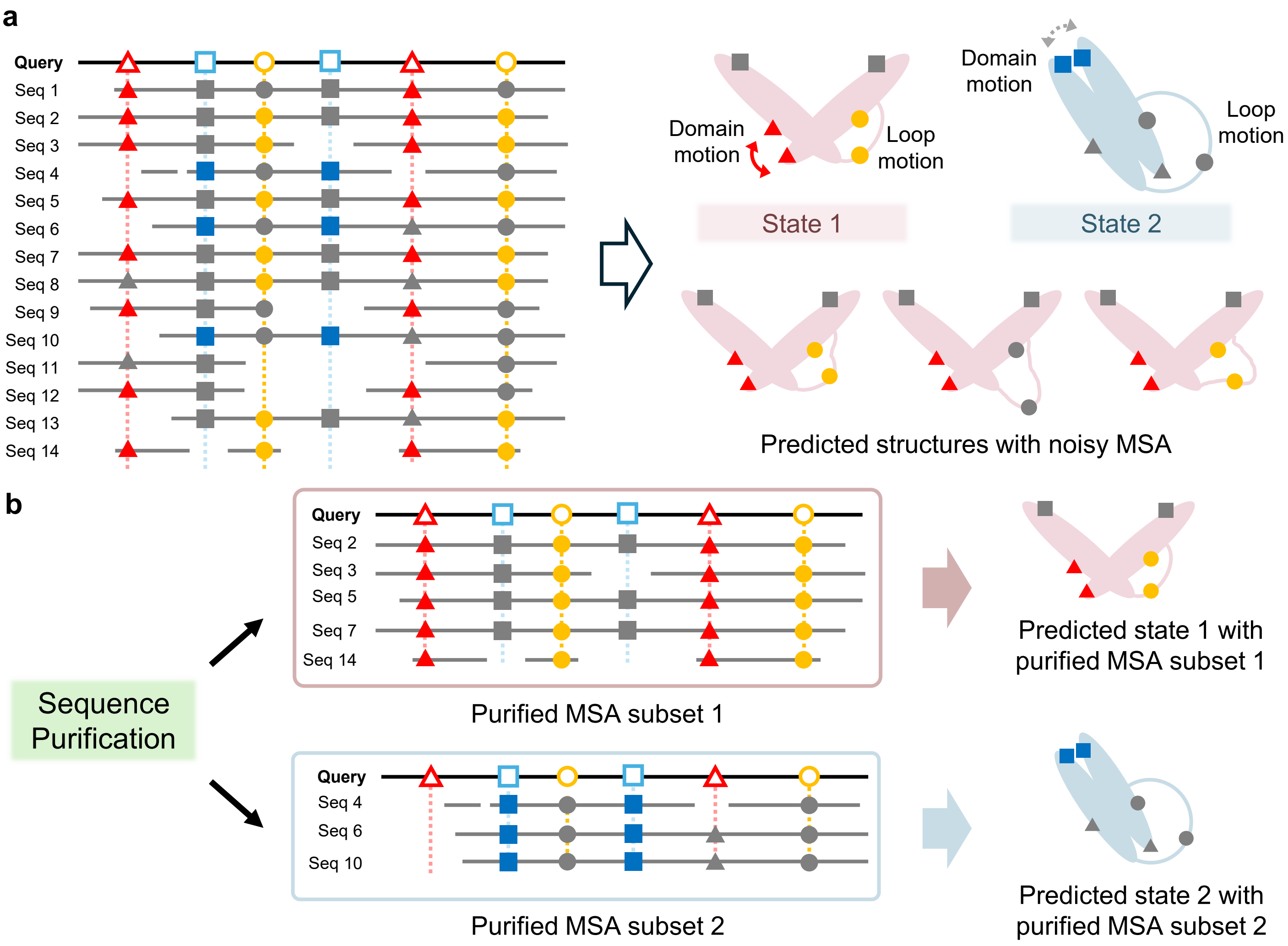
a
Query
Seq 1
Seq 2
Seq 3
Seq 4
Seq 5
Seq 6
Seq 7
Seq 8
Seq 9
Seq 10
Seq 11
Seq 12
Seq 13
Seq 14
Domain motion
Loop motion
State 1
Domain motion
Loop motion
State 2
Predicted structures with noisy MSA
b
Sequence Purification
Query
Seq 2
Seq 3
Seq 5
Seq 7
Seq 14
Purified MSA subset 1
Predicted state 1 with purified MSA subset 1
Query
Seq 4
Seq 6
Seq 10
Purified MSA subset 2
Predicted state 2 with purified MSA subset 2

# Figure 2

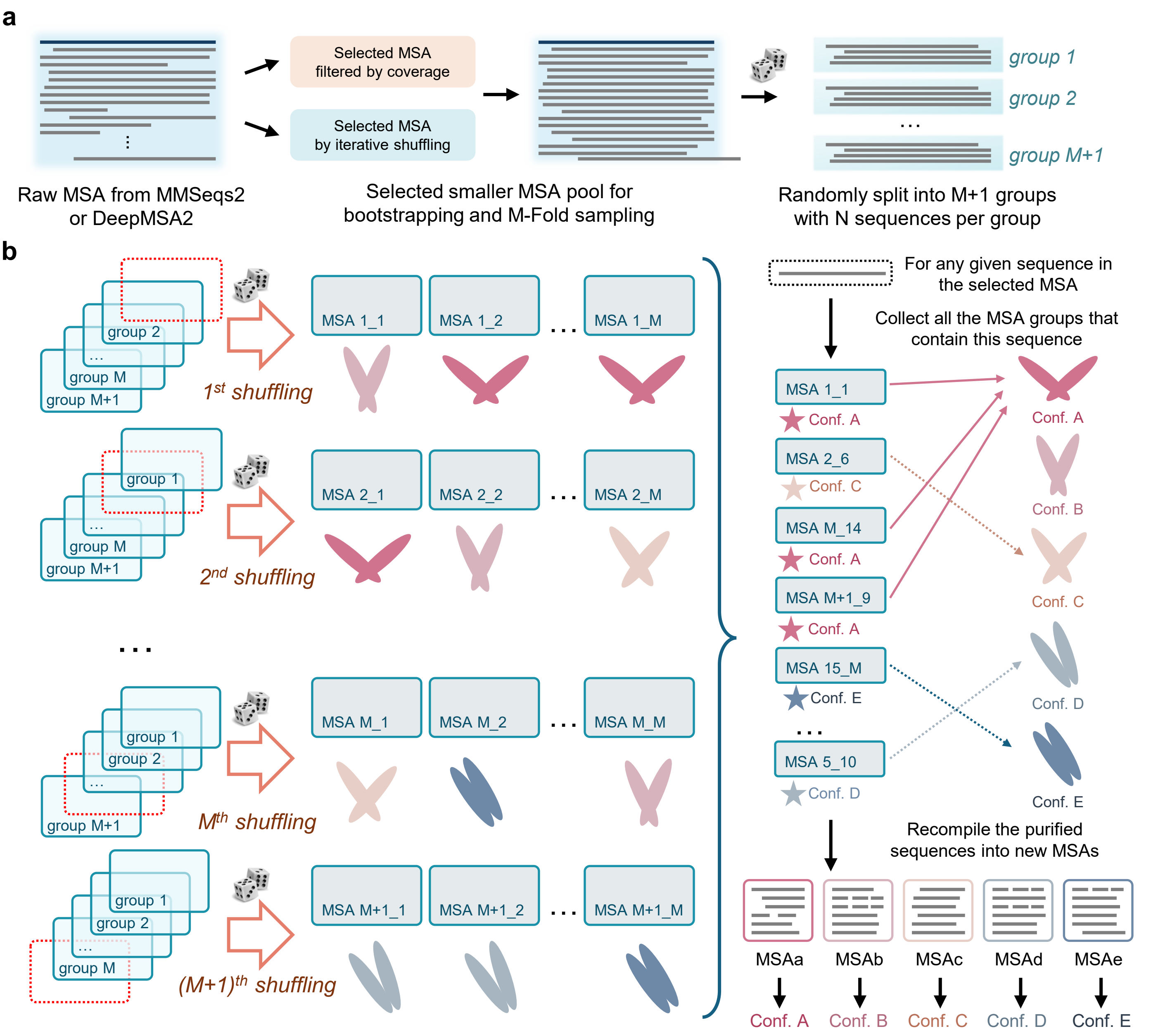
a
Selected MSA filtered by coverage
Selected MSA by iterative shuffling
group 1
group 2
group M+1
Raw MSA from MMSeqs2 or DeepMSA2
Selected smaller MSA pool for bootstrapping and M-Fold sampling
Randomly split into M+1 groups with N sequences per group
b
group 2
group M
group M+1
1st shuffling
MSA 1_1
MSA 1_2
MSA 1_M
group 1
group M
group M+1
2nd shuffling
MSA 2_1
MSA 2_2
MSA 2_M
group 1
group 2
group M+1
Mth shuffling
MSA M_1
MSA M_2
MSA M_M
group 1
group 2
group M
(M+1)th shuffling
MSA M+1_1
MSA M+1_2
MSA M+1_M
For any given sequence in the selected MSA
Collect all the MSA groups that contain this sequence
MSA 1_1
Conf. A
MSA 2_6
Conf. C
MSA M_14
Conf. A
MSA M+1_9
Conf. A
MSA 15_M
Conf. E
MSA 5_10
Conf. D
Conf. A
Conf. B
Conf. C
Conf. D
Conf. E
Recompile the purified sequences into new MSAs
MSAa
MSAb
MSAc
MSAd
MSAe
Conf. A
Conf. B
Conf. C
Conf. D
Conf. E

# Figure 3

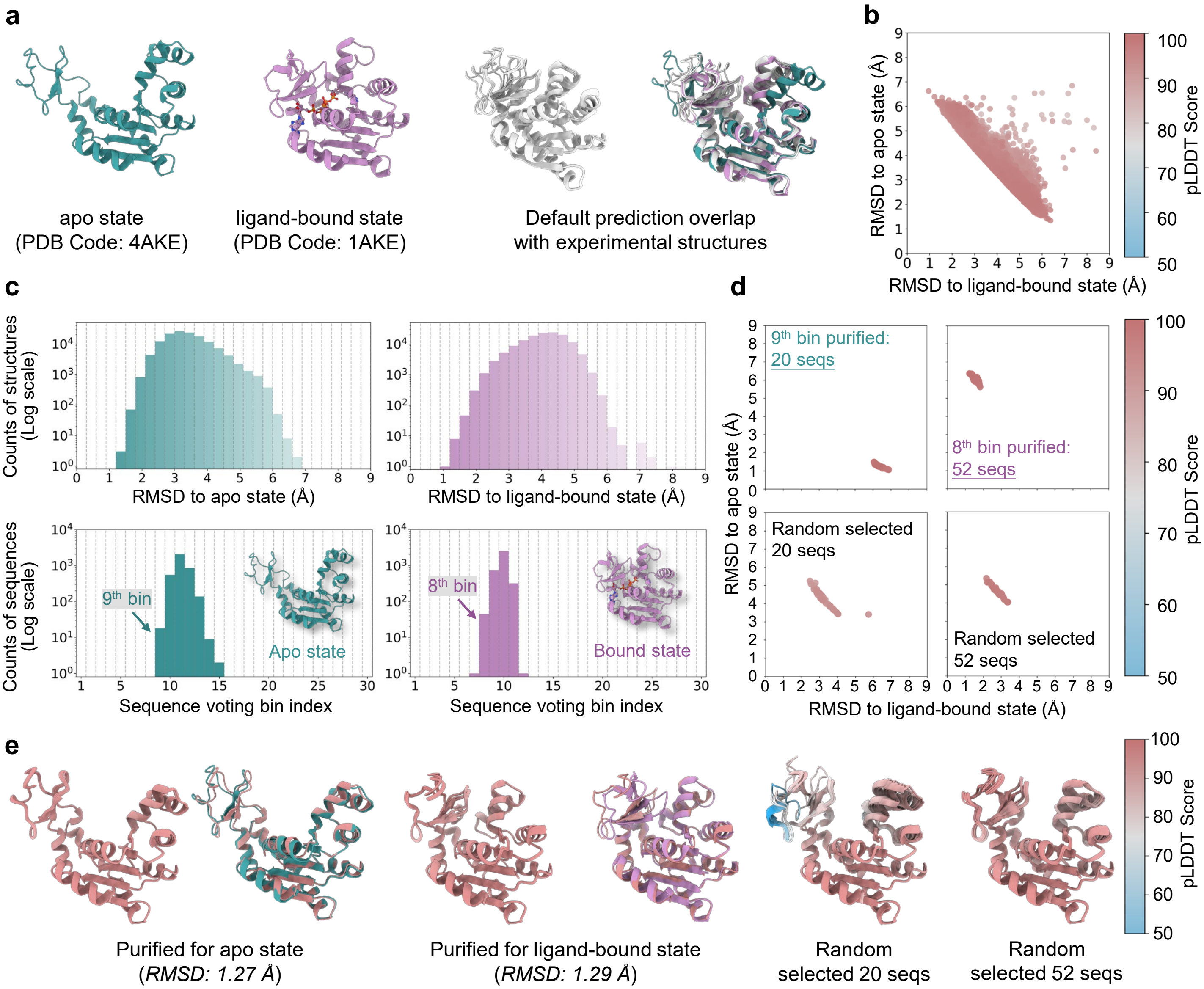

# Figure 4

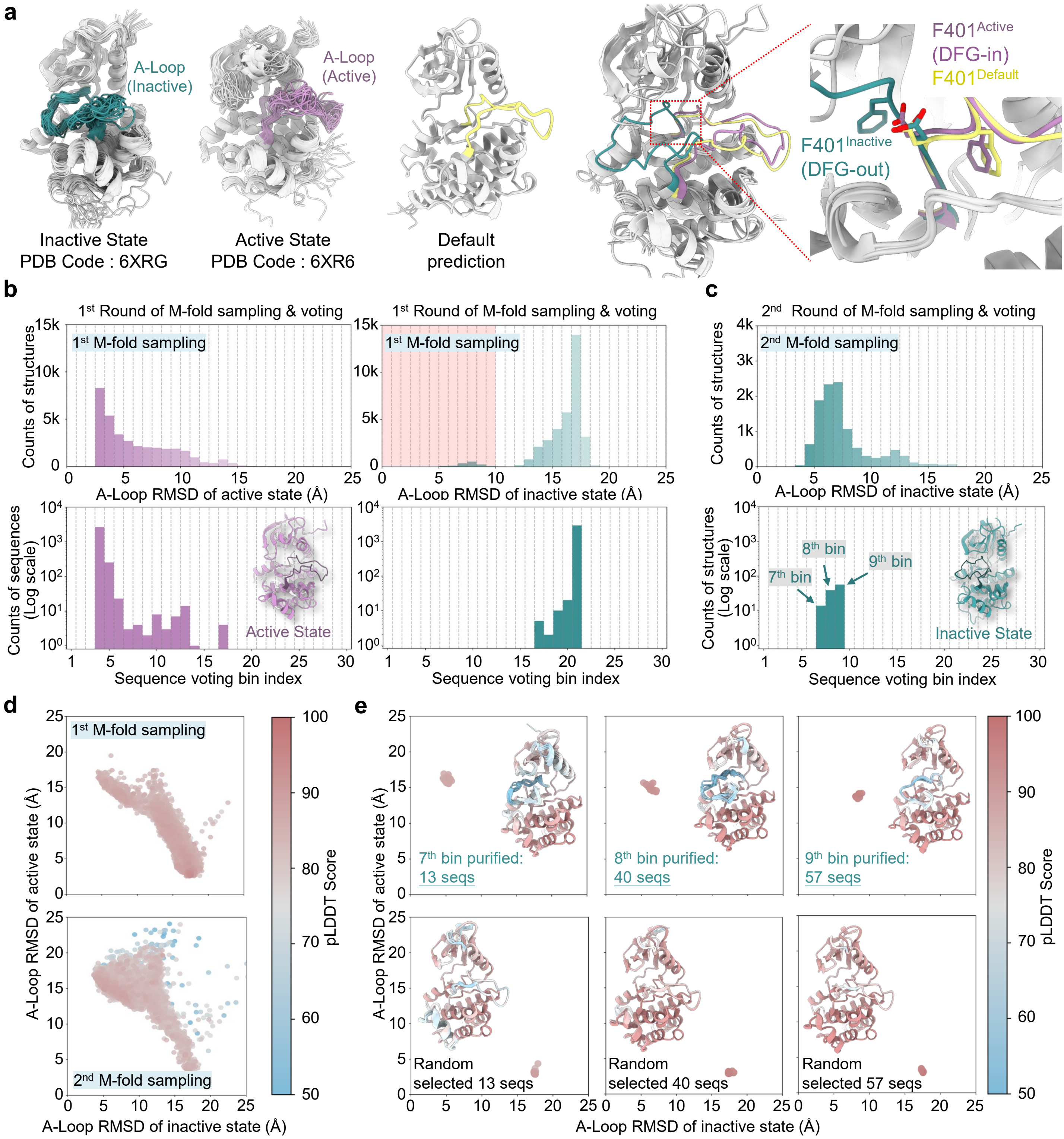
a
A-Loop (Inactive)
A-Loop (Active)
Inactive State PDB Code : 6XRG
Active State PDB Code : 6XR6
Default prediction
F401Active (DFG-in)
F401Default
F401Inactive (DFG-out)
b
1st Round of M-fold sampling & voting
1st M-fold sampling
Counts of structures
A-Loop RMSD of active state (Å)
A-Loop RMSD of inactive state (Å)
Counts of sequences (Log scale)
Active State
Sequence voting bin index
c
2nd Round of M-fold sampling & voting
2nd M-fold sampling
Counts of structures (Log scale)
7th bin
8th bin
9th bin
Inactive State
d
2nd M-fold sampling
A-Loop RMSD of active state (Å)
pLDDT Score
e
7th bin purified: 13 seqs
8th bin purified: 40 seqs
9th bin purified: 57 seqs
Random selected 13 seqs
Random selected 40 seqs
Random selected 57 seqs

# Figure 5

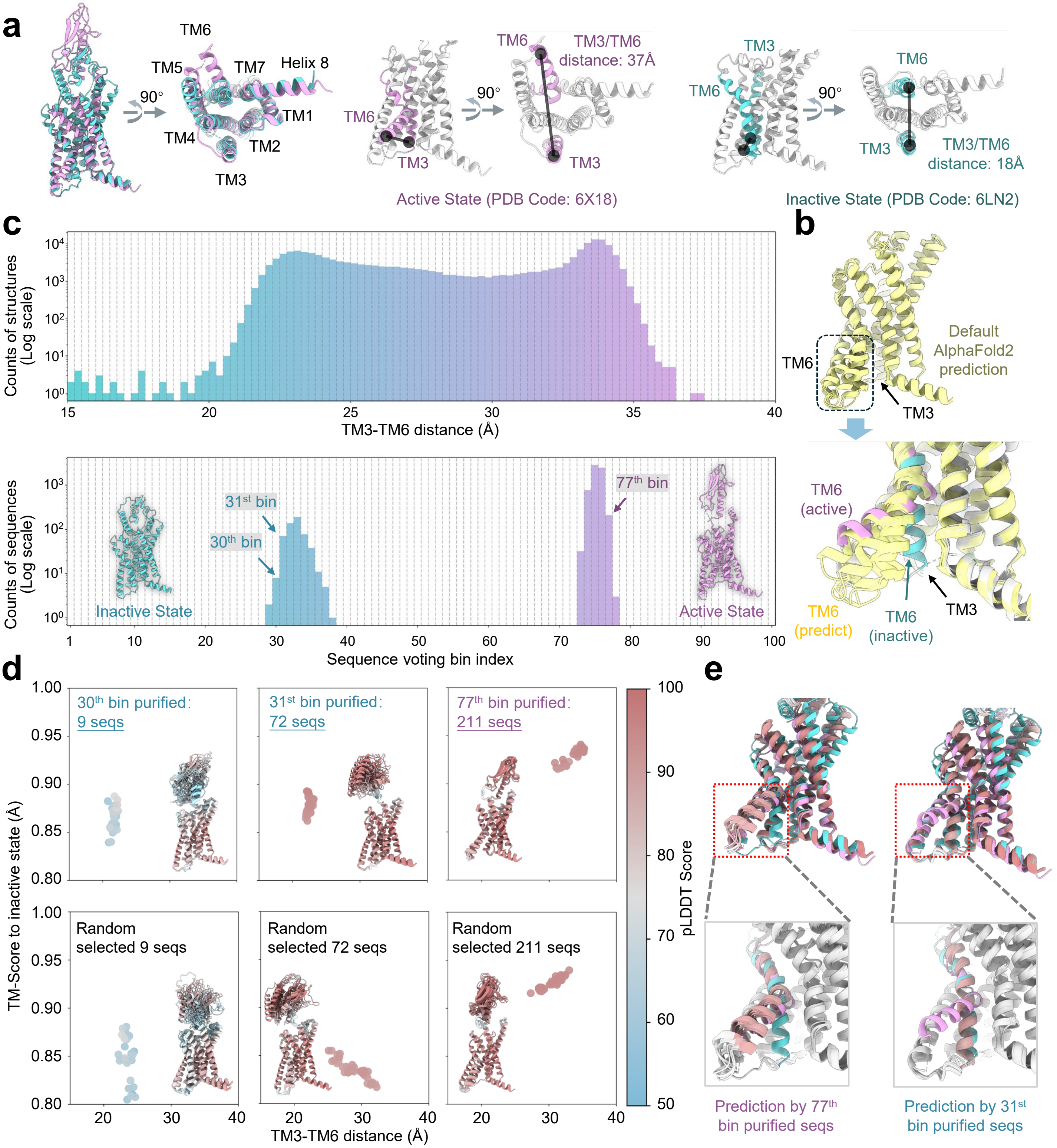
a
TM6
TM5
TM7
Helix 8
90°
TM1
TM4
TM2
TM3
TM6
TM3/TM6 distance: 37Å
90°
TM6
TM3
TM3
Active State (PDB Code: 6X18)
TM3
TM6
TM6
90°
TM3
TM3/TM6 distance: 18Å
Inactive State (PDB Code: 6LN2)
c
Counts of structures (Log scale)
TM3-TM6 distance (Å)
b
Default AlphaFold2 prediction
TM6
TM3
TM6 (active)
TM6 (predict)
TM6 (inactive)
TM3
Counts of sequences (Log scale)
31st bin
30th bin
77th bin
Inactive State
Active State
Sequence voting bin index
d
30th bin purified: 9 seqs
31st bin purified: 72 seqs
77th bin purified: 211 seqs
Random selected 9 seqs
Random selected 72 seqs
Random selected 211 seqs
TM-Score to inactive state (Å)
TM3-TM6 distance (Å)
pLDDT Score
e
Prediction by 77th bin purified seqs
Prediction by 31st bin purified seqs

# Figure 6

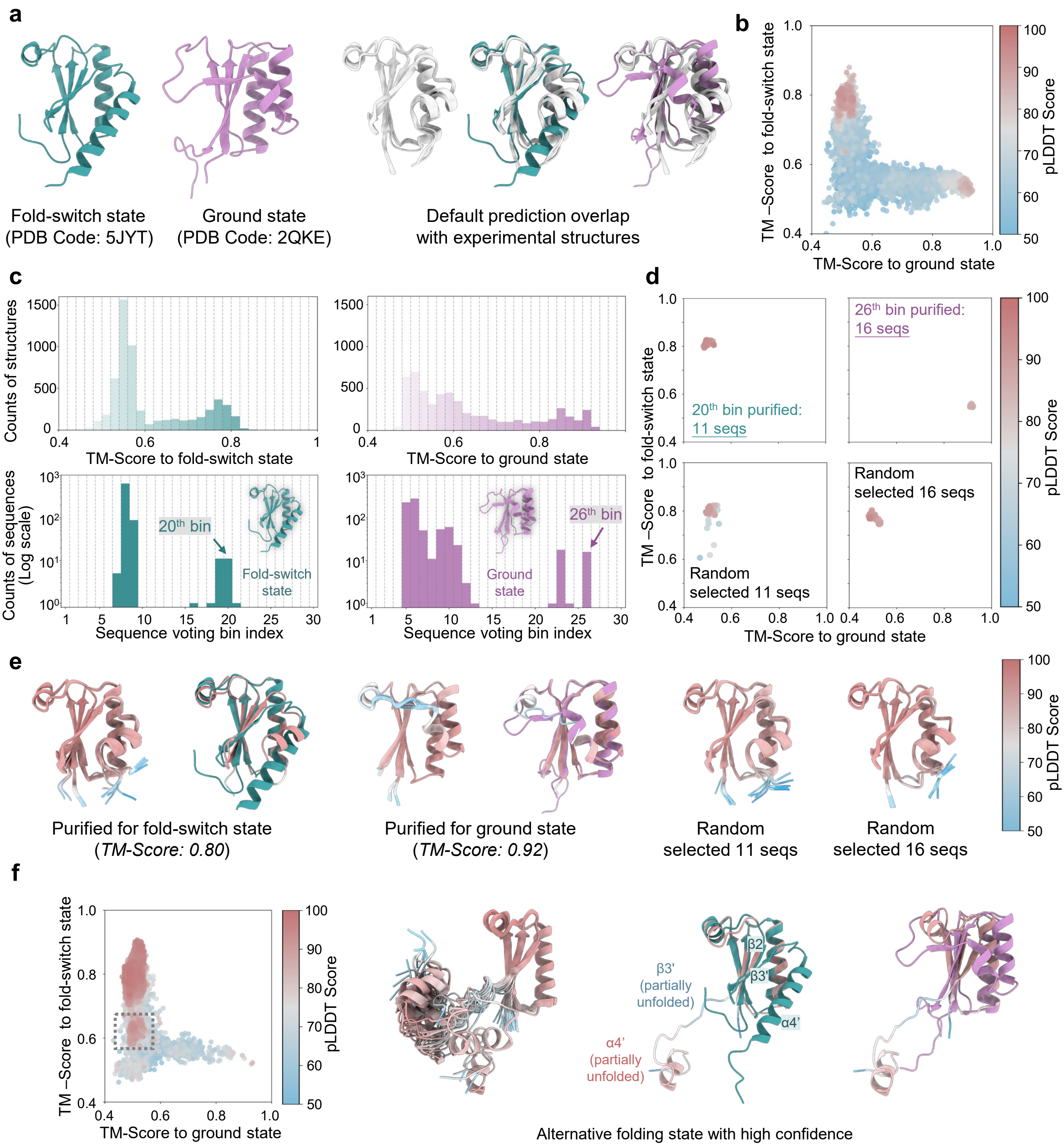

# Figure 7

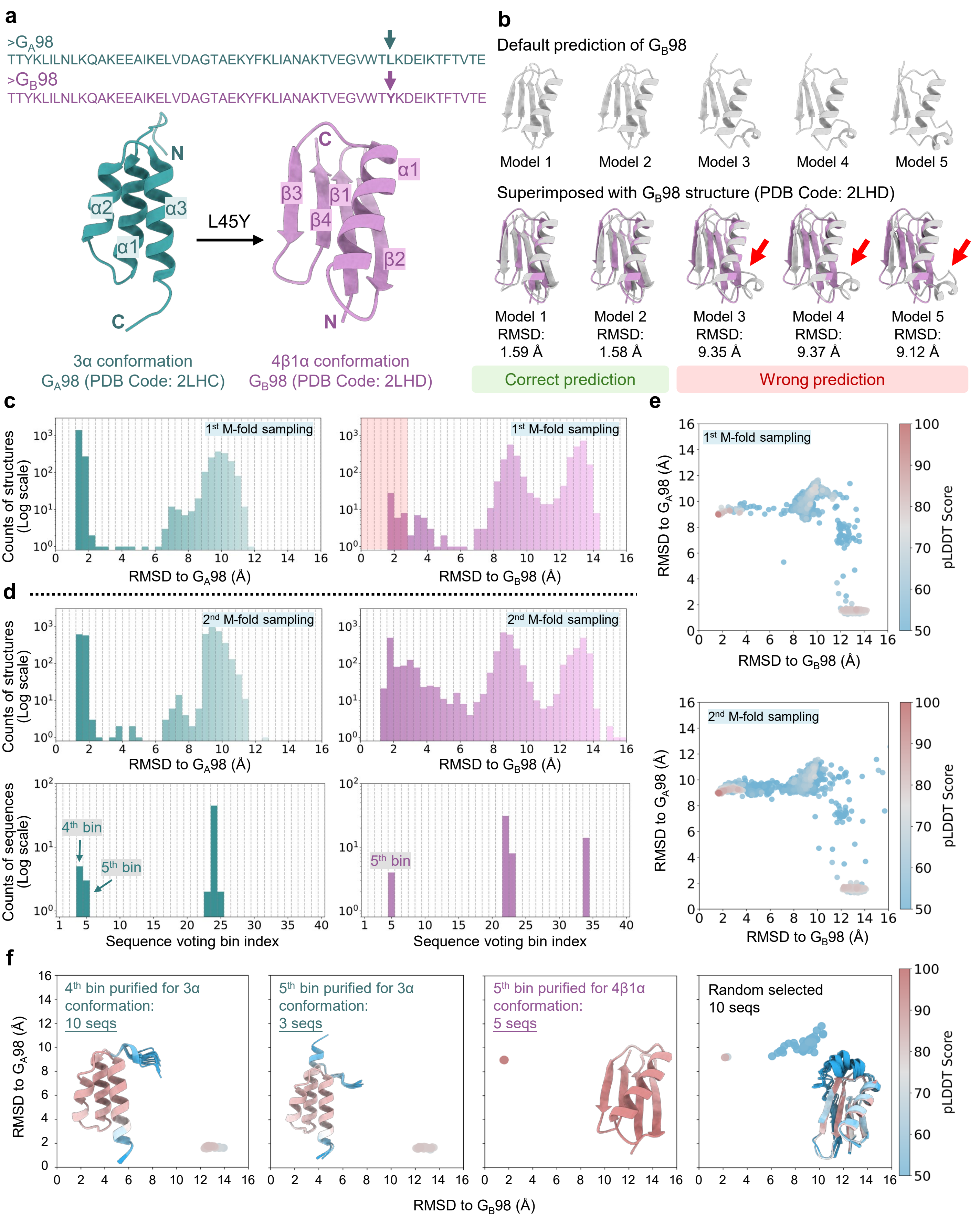
a
>G$_A$98
TTYKLILNLKQAKEEAIKELVDAGTAEKYFKLIANAKTVEGVWTLKDEIKTFTVTE
>G$_B$98
TTYKLILNLKQAKEEAIKELVDAGTAEKYFKLIANAKTVEGVWTYKDEIKTFTVTE
L45Y
3α conformation
G$_A$98 (PDB Code: 2LHC)
4β1α conformation
G$_B$98 (PDB Code: 2LHD)
b
Default prediction of G$_B$98
Model 1
Model 2
Model 3
Model 4
Model 5
Superimposed with G$_B$98 structure (PDB Code: 2LHD)
Model 1 RMSD: 1.59 Å
Model 2 RMSD: 1.58 Å
Model 3 RMSD: 9.35 Å
Model 4 RMSD: 9.37 Å
Model 5 RMSD: 9.12 Å
Correct prediction
Wrong prediction
c
1st M-fold sampling
Counts of structures (Log scale)
RMSD to G$_A$98 (Å)
RMSD to G$_B$98 (Å)
d
2nd M-fold sampling
Counts of sequences (Log scale)
4th bin
5th bin
Sequence voting bin index
e
1st M-fold sampling
2nd M-fold sampling
RMSD to G$_A$98 (Å)
RMSD to G$_B$98 (Å)
pLDDT Score
f
4th bin purified for 3α conformation: 10 seqs
5th bin purified for 3α conformation: 3 seqs
5th bin purified for 4β1α conformation: 5 seqs
Random selected 10 seqs
RMSD to G$_A$98 (Å)
RMSD to G$_B$98 (Å)
pLDDT Score

# Figure 8

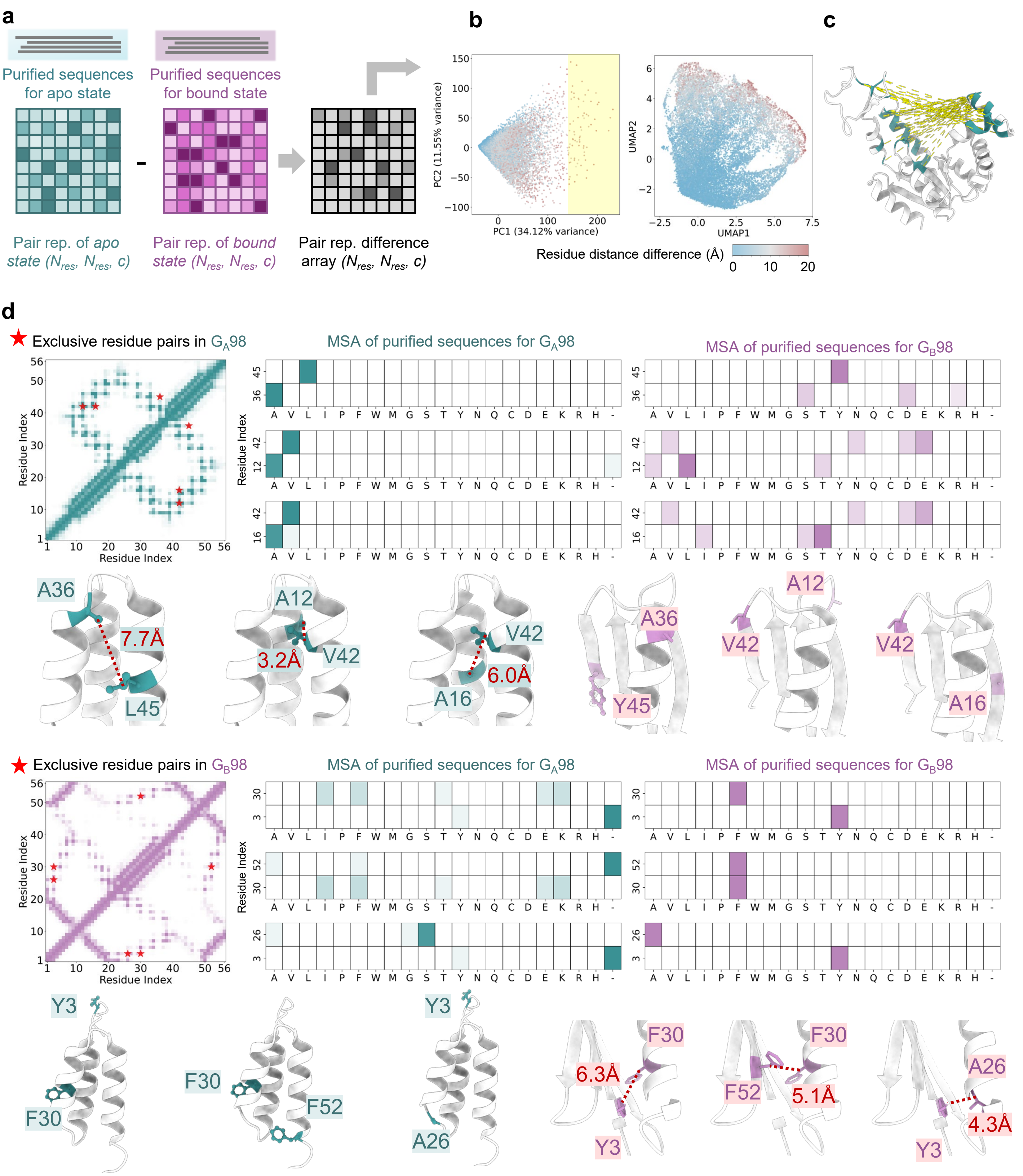

a
Purified sequences for apo state
Purified sequences for bound state
Pair rep. of apo state (N_res, N_res, c)
Pair rep. of bound state (N_res, N_res, c)
Pair rep. difference array (N_res, N_res, c)
b
PC1 (34.12% variance)
PC2 (11.55% variance)
UMAP1
UMAP2
Residue distance difference (Å)
c
d
Exclusive residue pairs in GA98
MSA of purified sequences for GA98
MSA of purified sequences for GB98
Residue Index
A V L I P F W M G S T Y N Q C D E K R H -
A36
7.7Å
L45
A12
3.2Å
V42
6.0Å
A16
Y45
Exclusive residue pairs in GB98
Y3
F30
F52
A26
6.3Å
5.1Å
4.3Å

# Extended Data Figure. 1

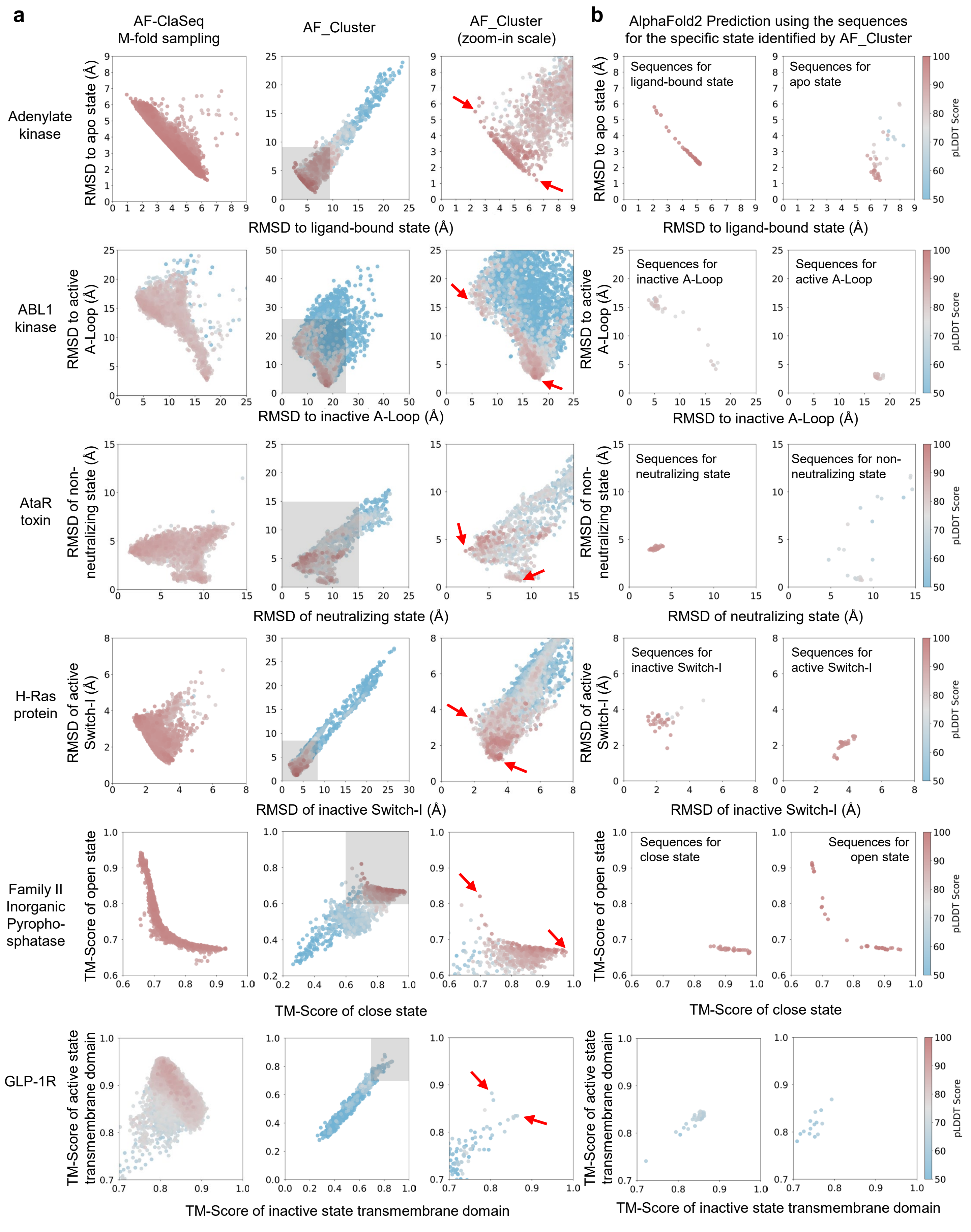

# Extended Data Figure. 1 Continued

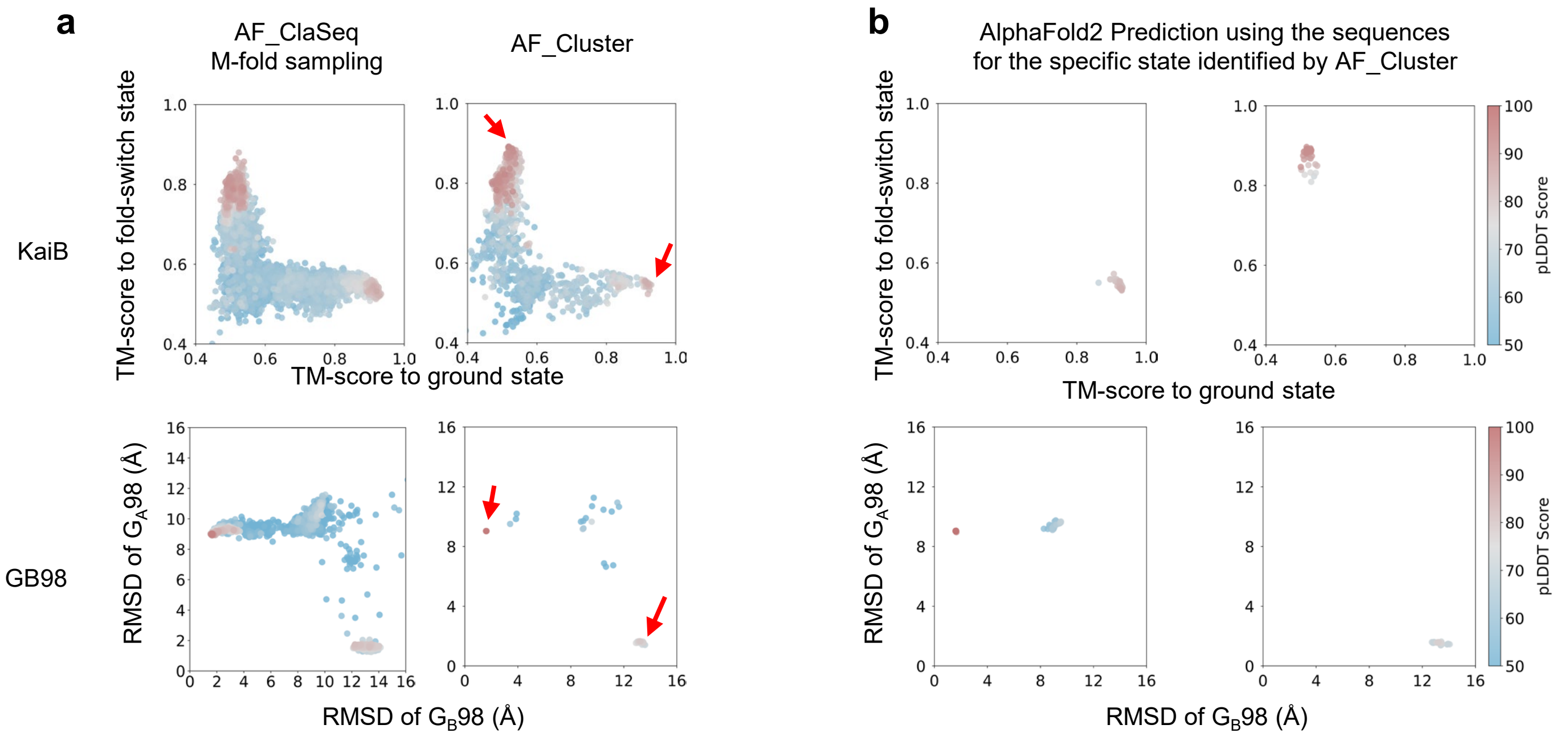

# Extended Data Figure. 2

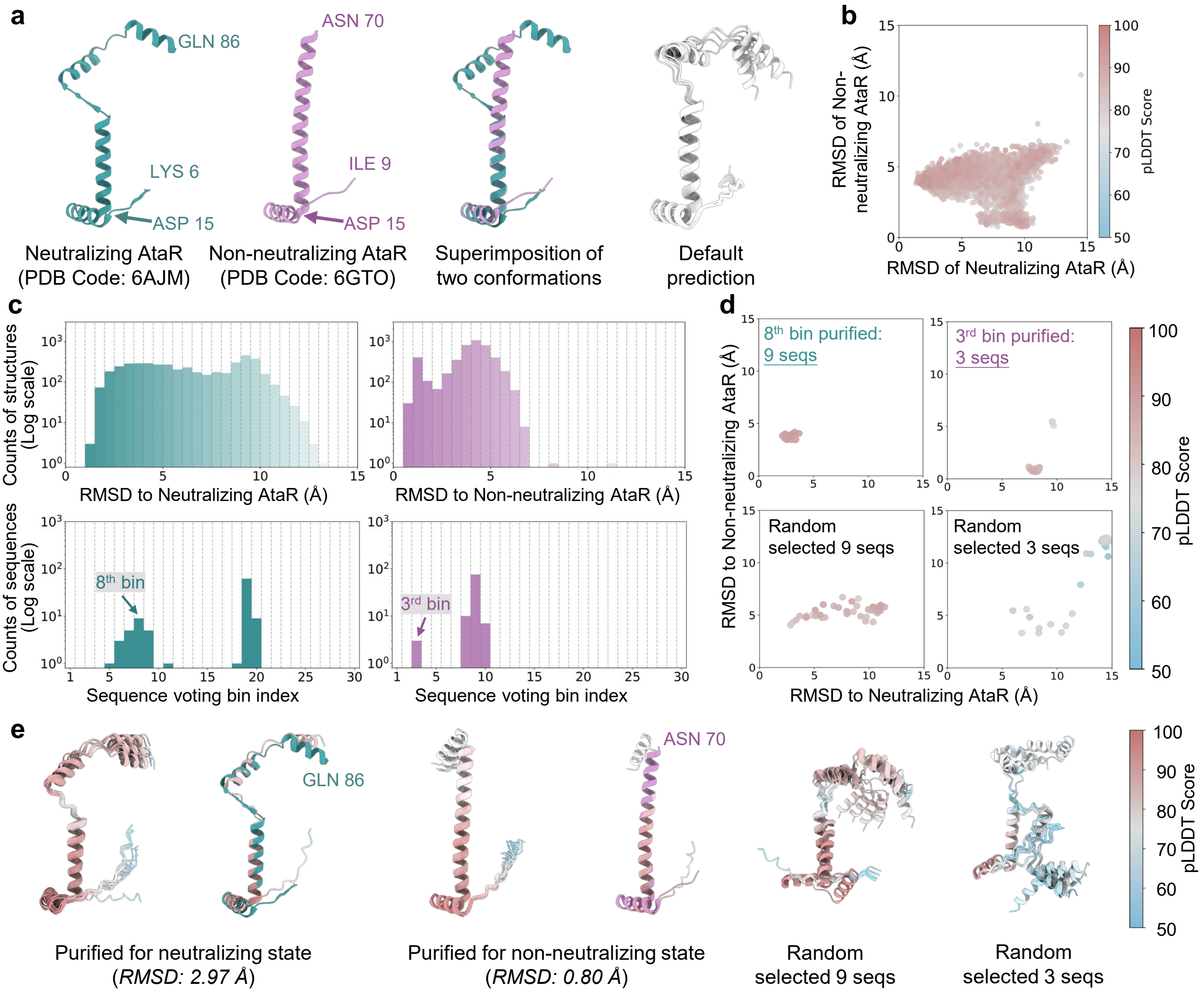

# Extended Data Figure. 3

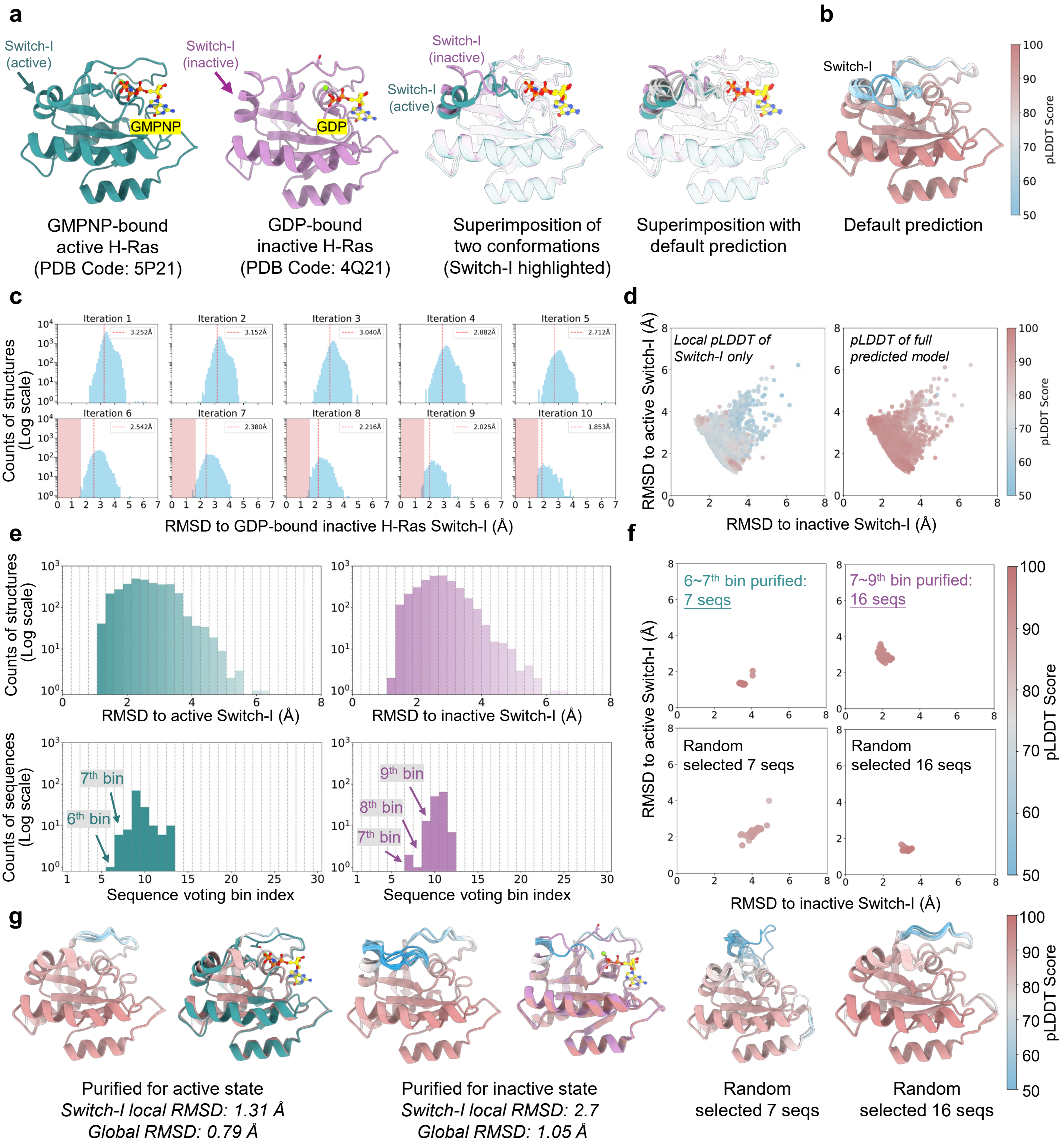

# Extended Data Figure. 4

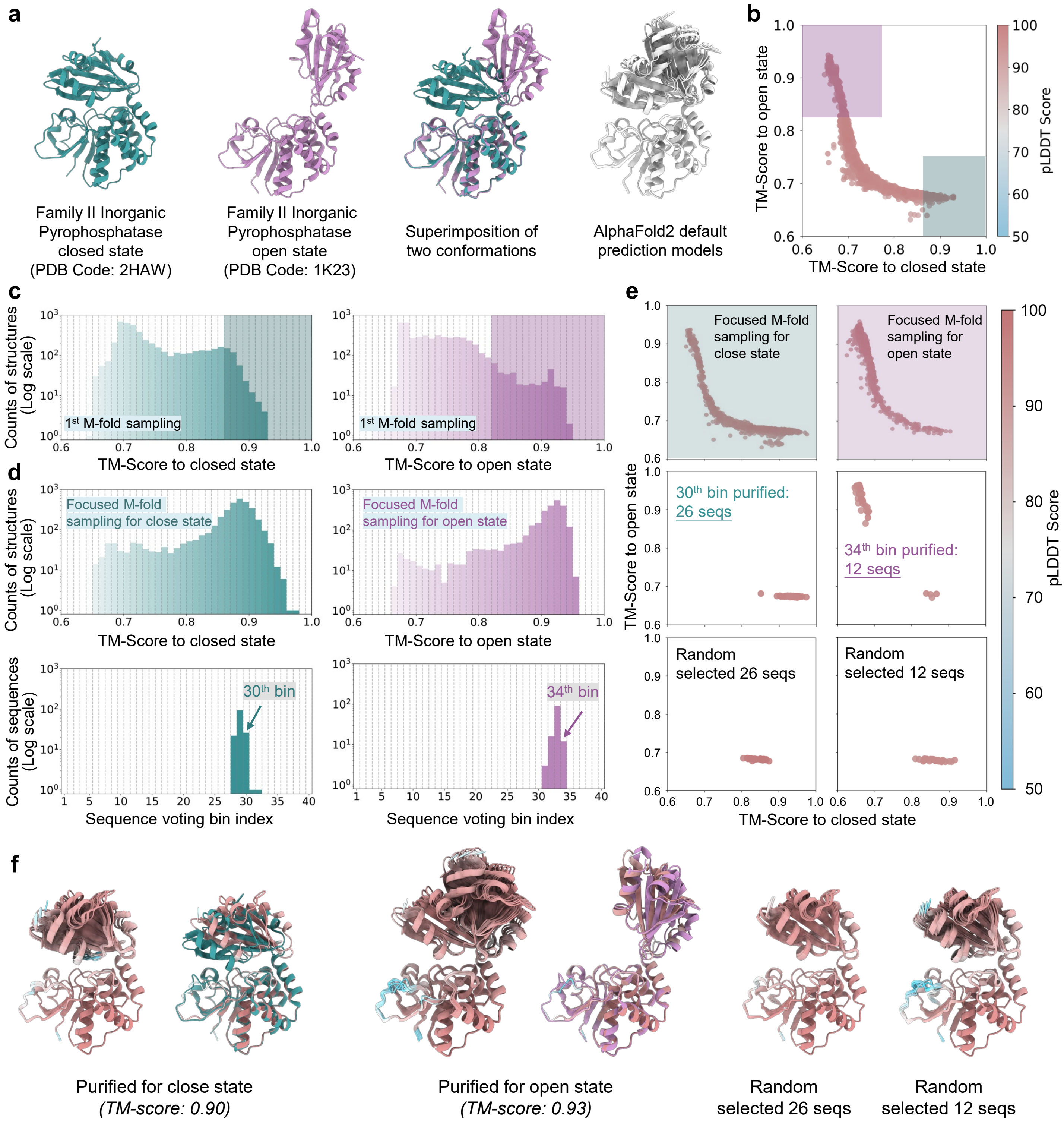

# Extended Data Figure. 5

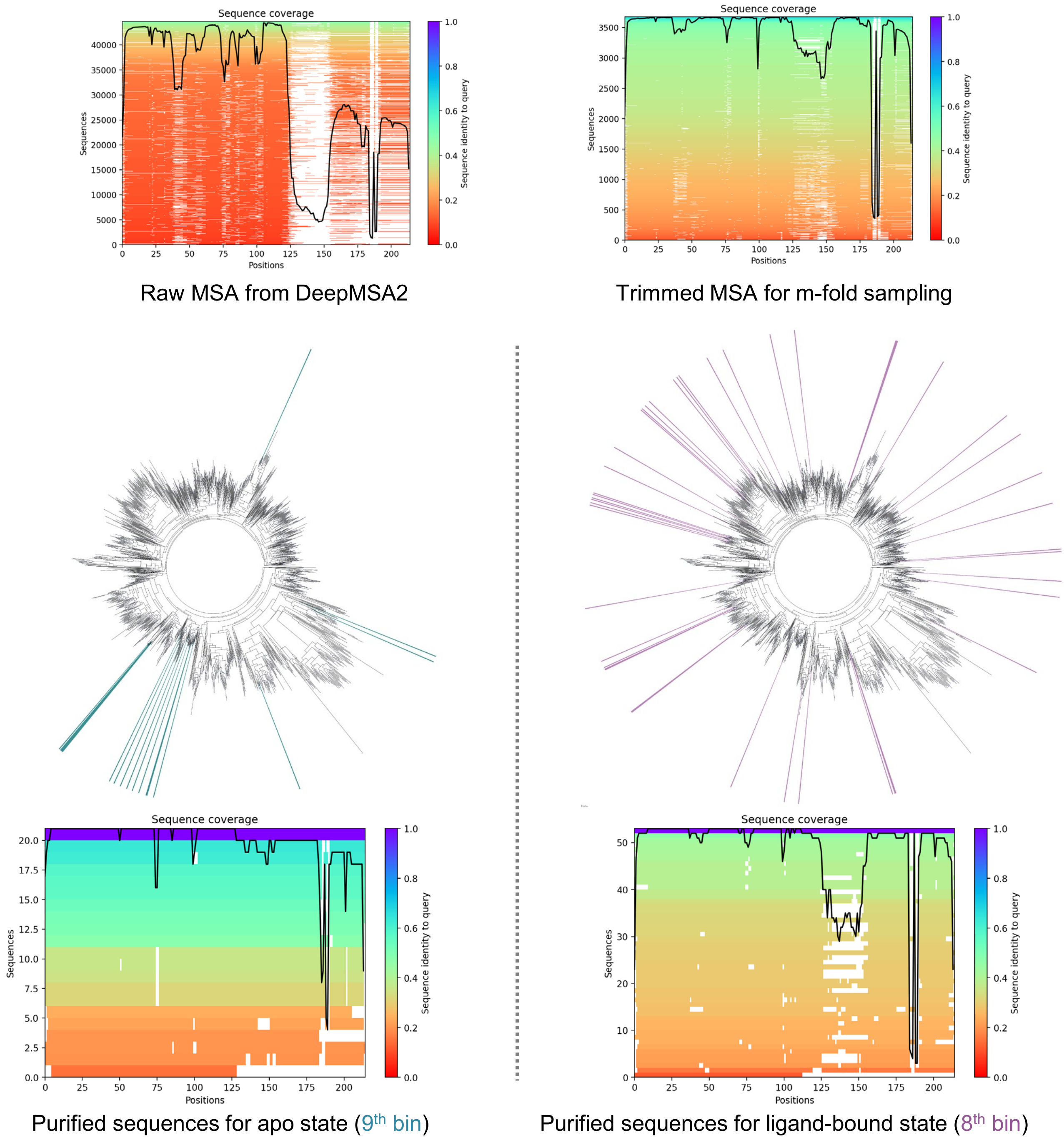

# Extended Data Figure. 6

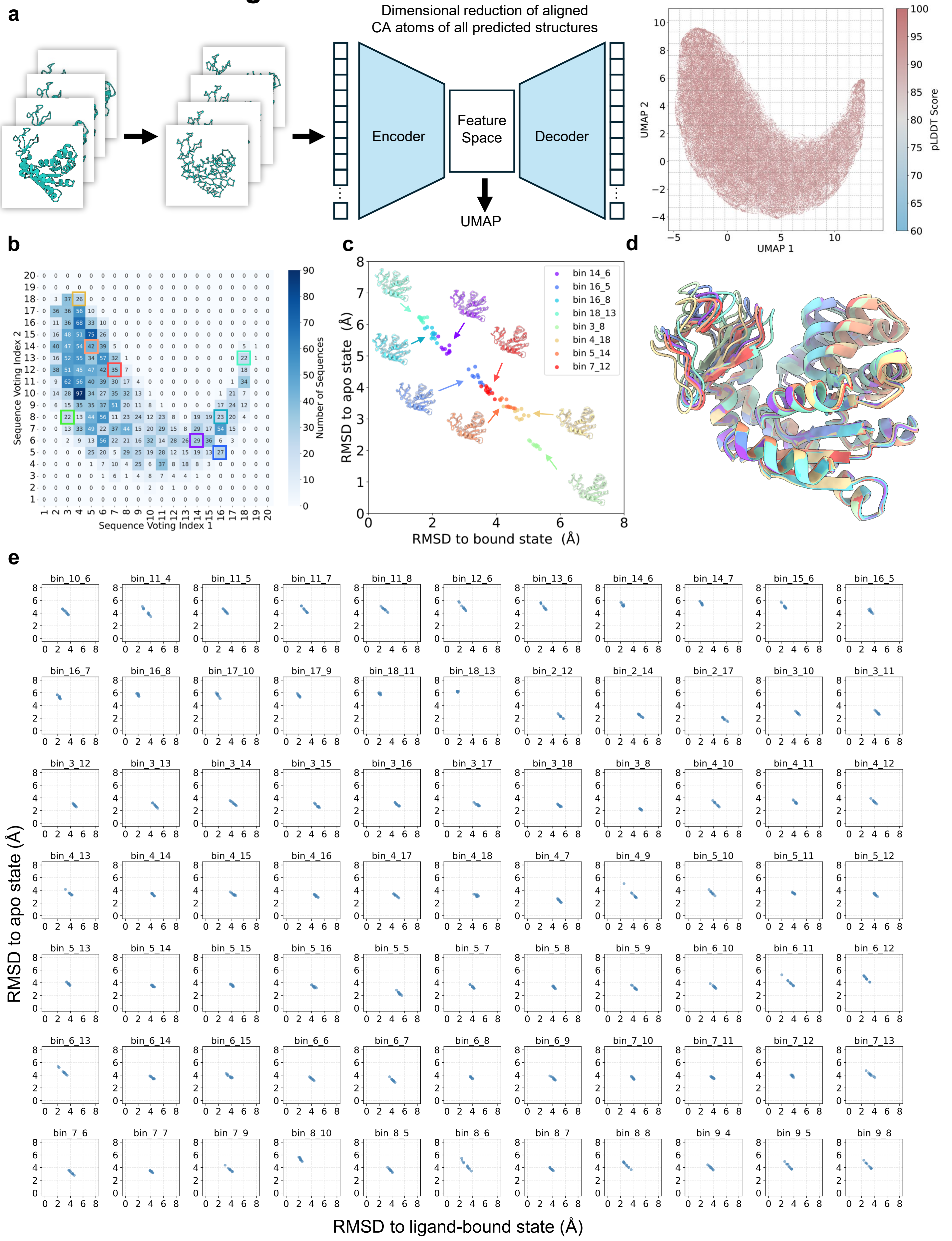

# Extended Data Figure. 7

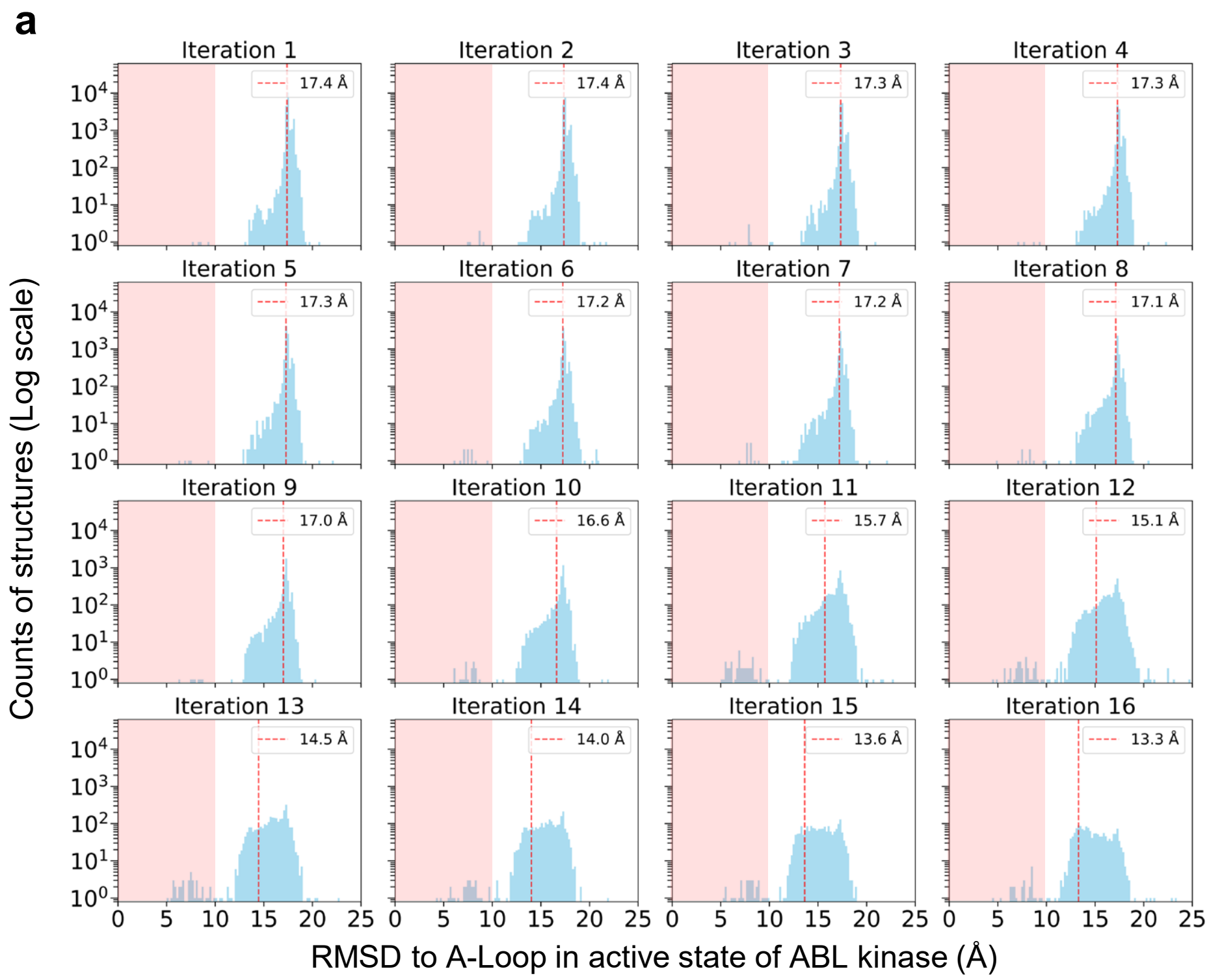

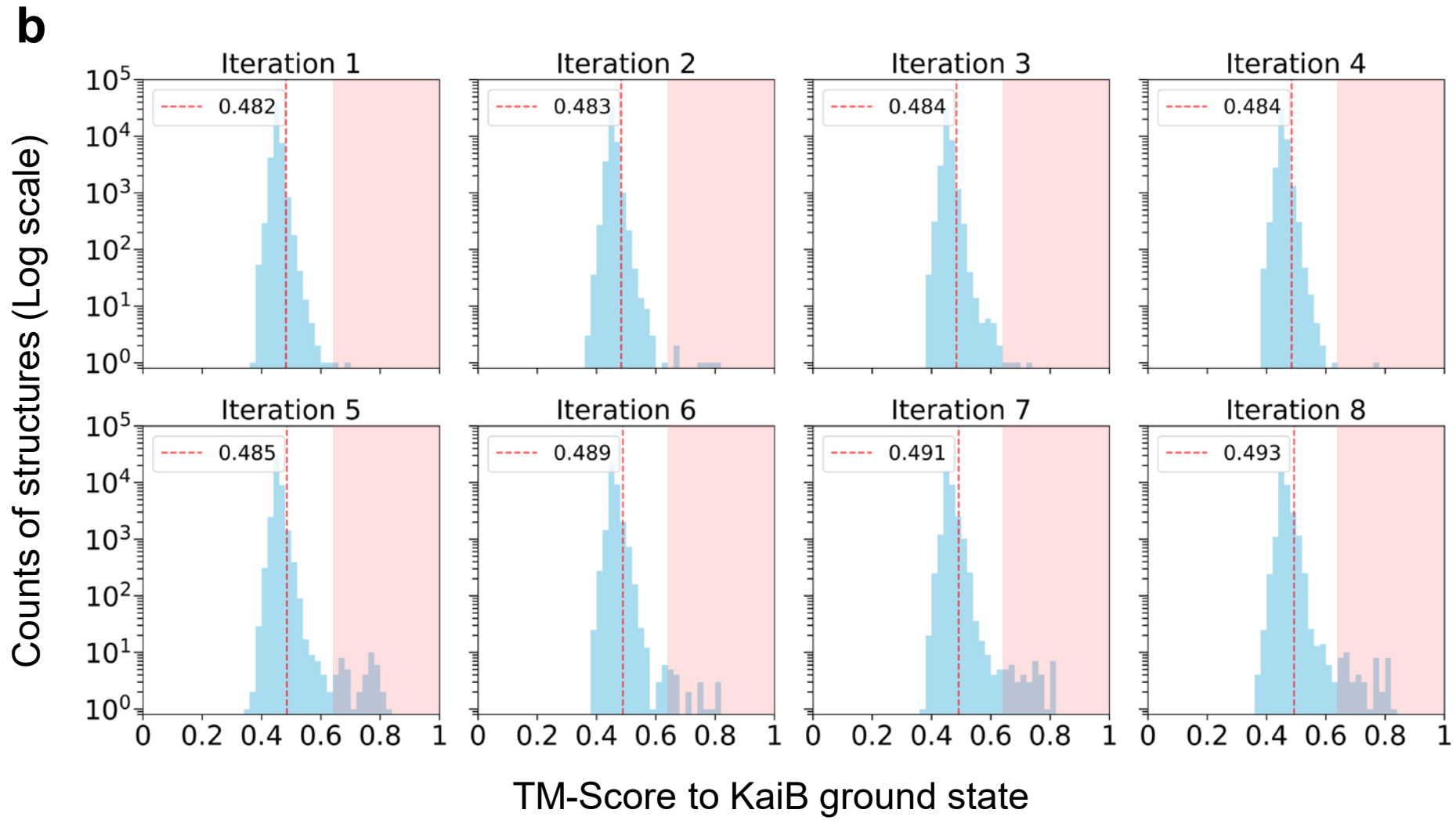

# Extended Data Figure. 8

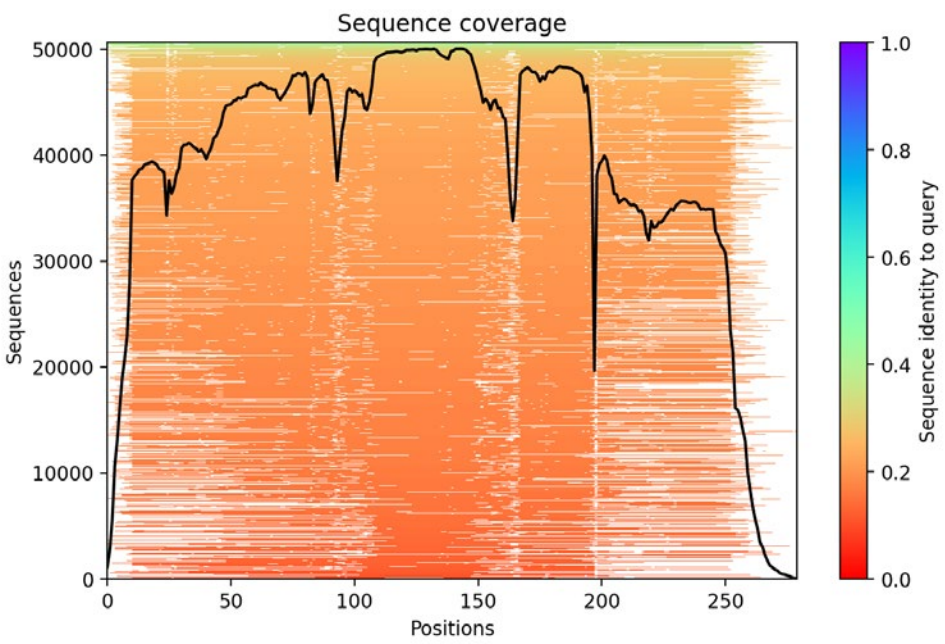

Raw MSA from DeepMSA2

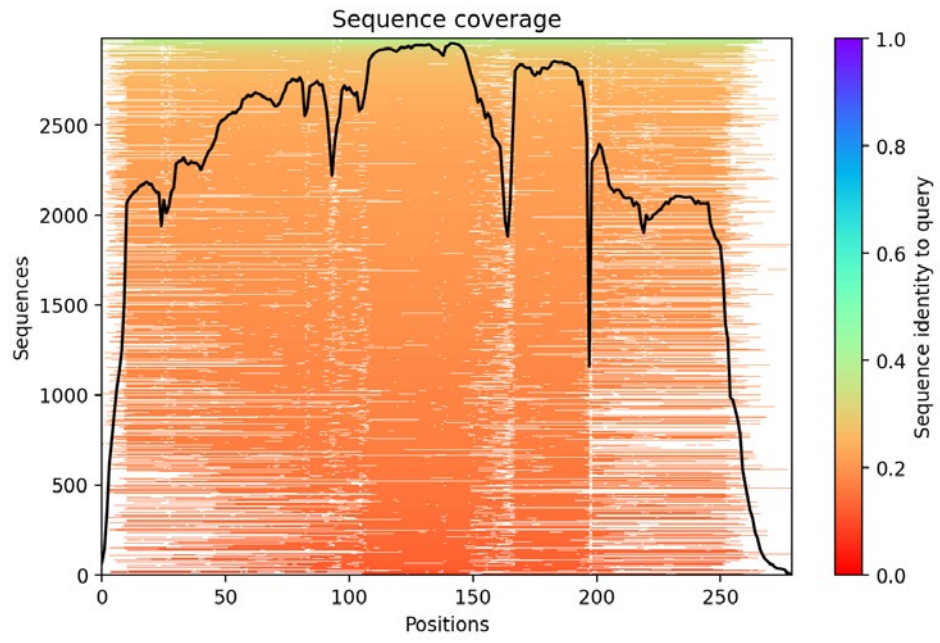

Enriched MSA for 1st round of M-fold sampling

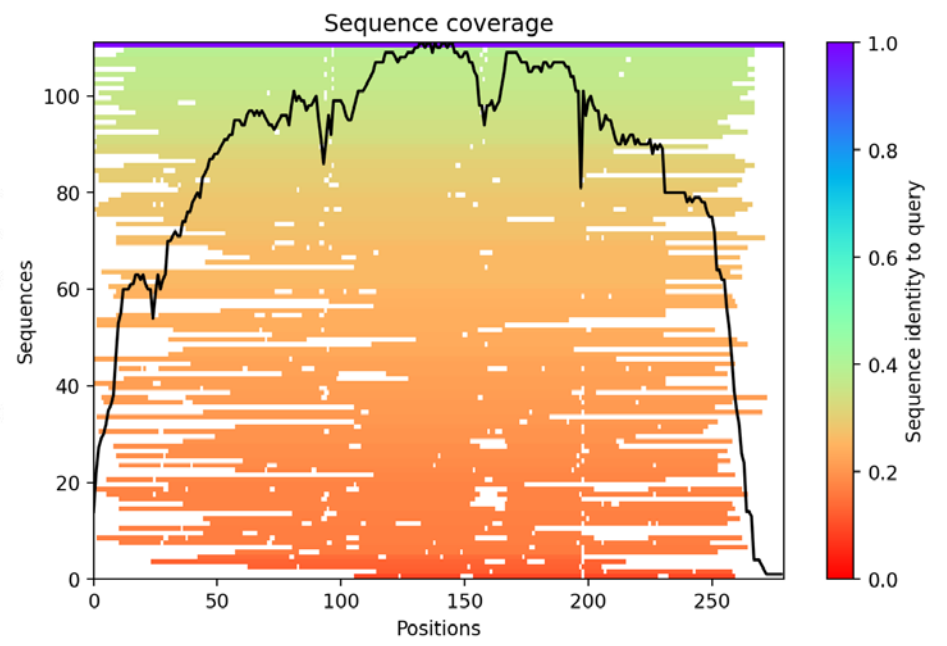

Enriched MSA for 2nd round of M-fold sampling

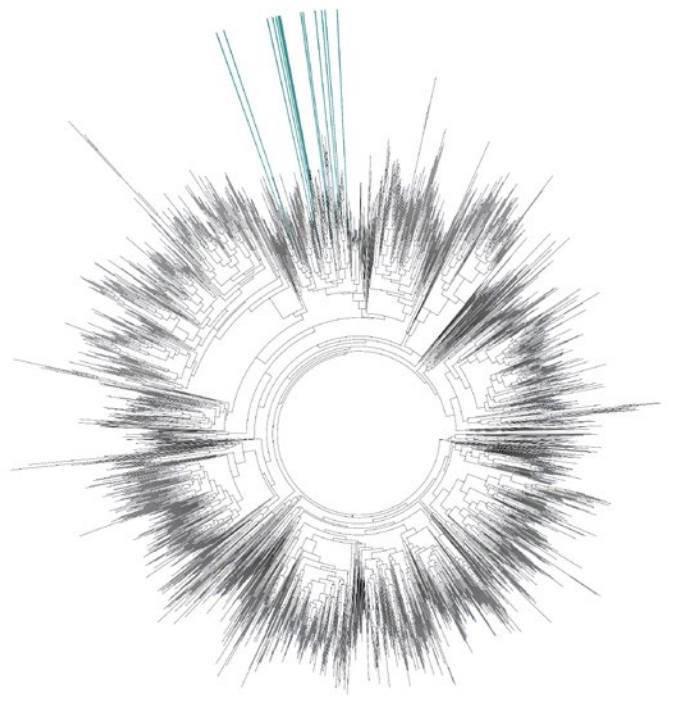

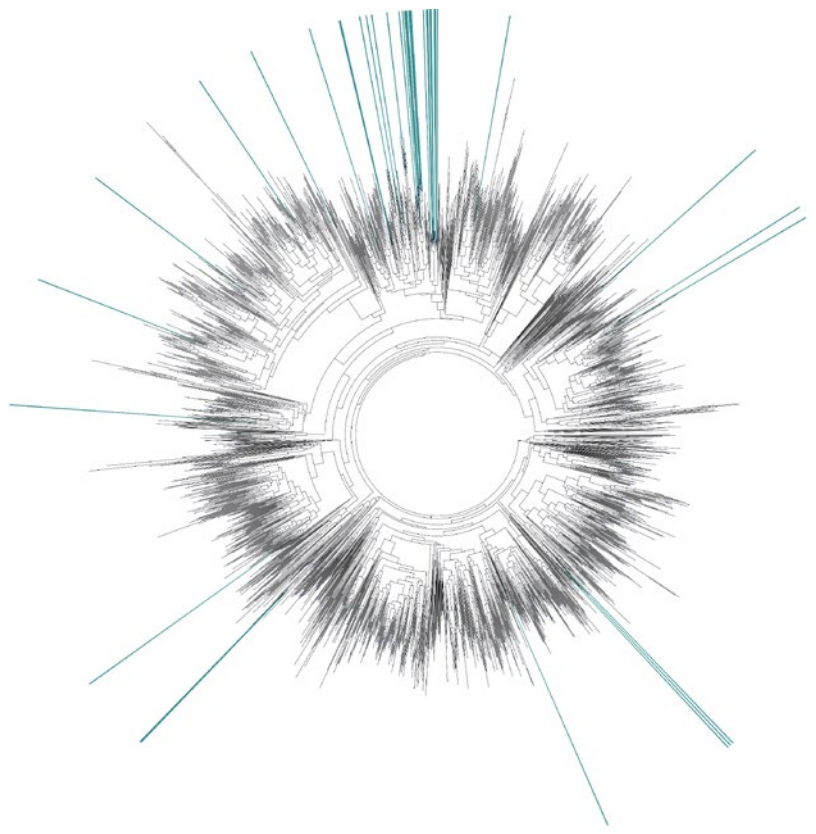

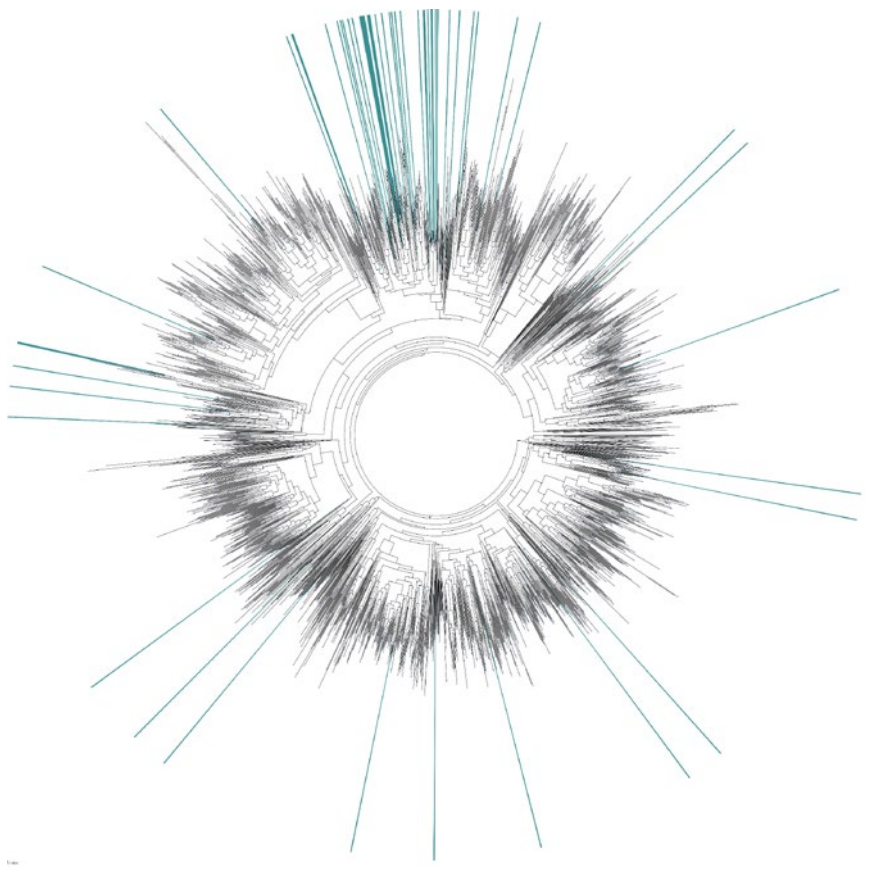

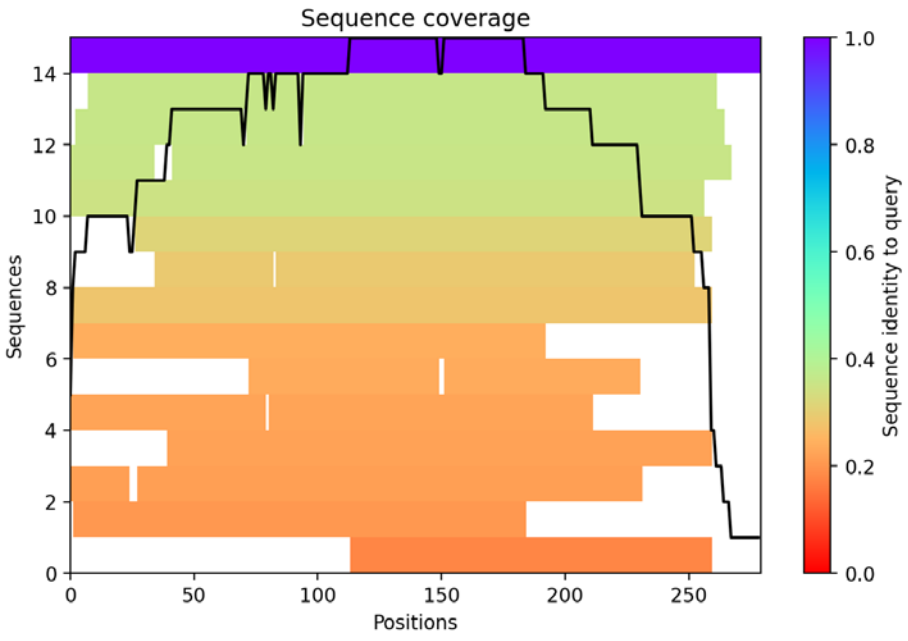

Purified sequences for inactive DFG-out state (7th bin)

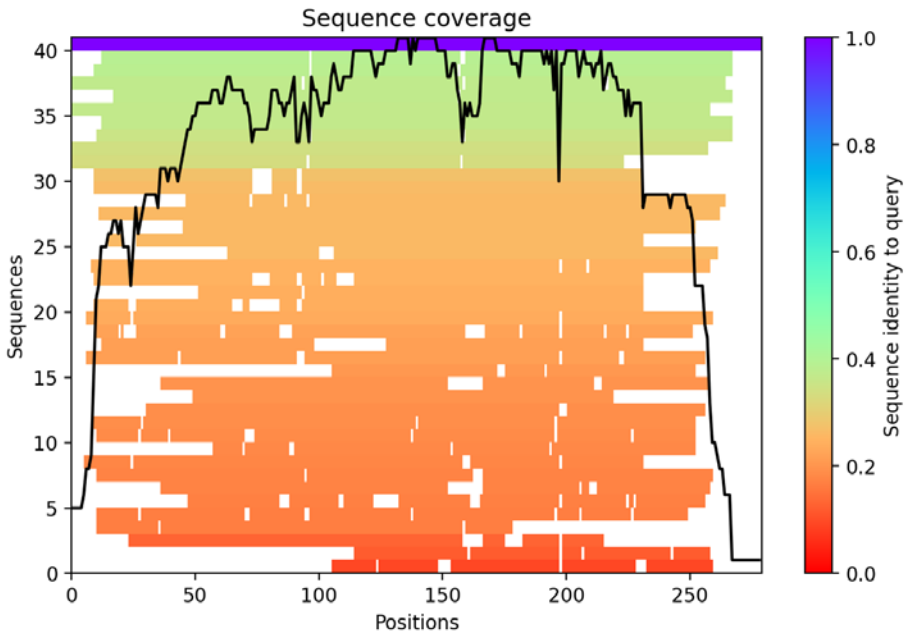

Purified sequences for inactive DFG-out state (8th bin)

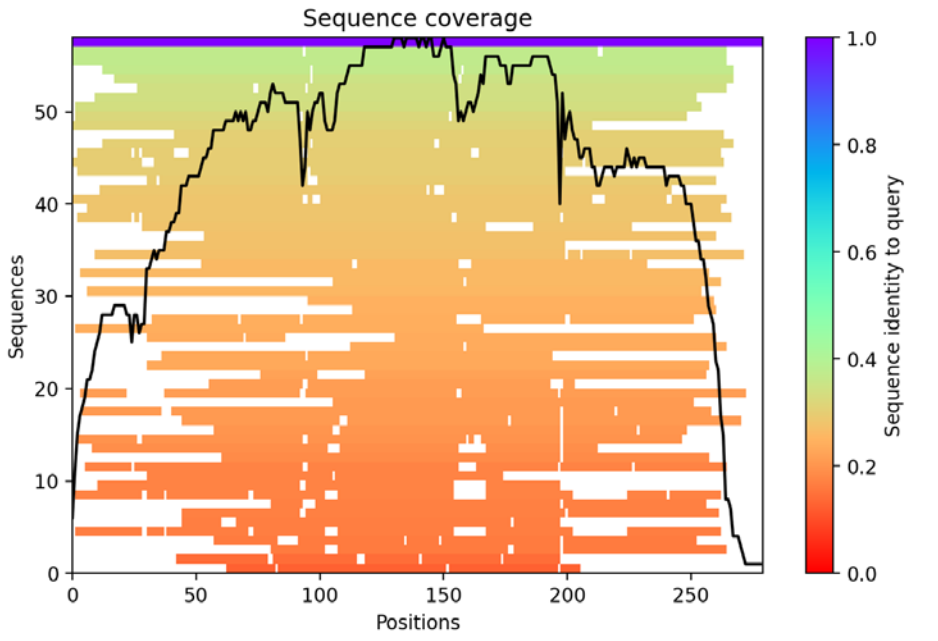

Purified sequences for inactive DFG-out state (9th bin)

# Extended Data Figure. 9

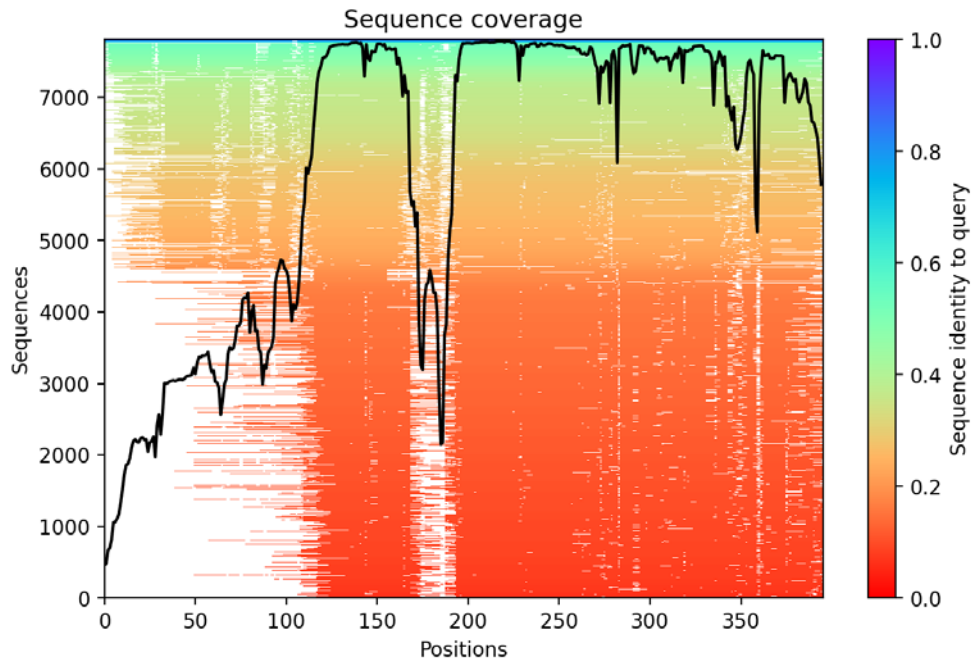

Raw MSA from DeepMSA2

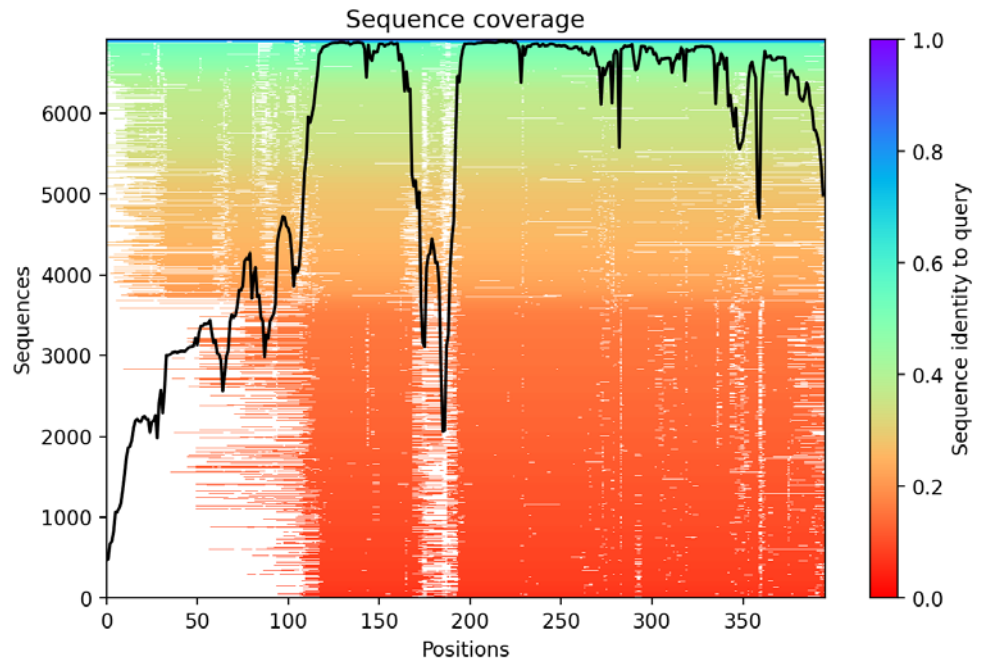

Trimmed MSA for m-fold sampling

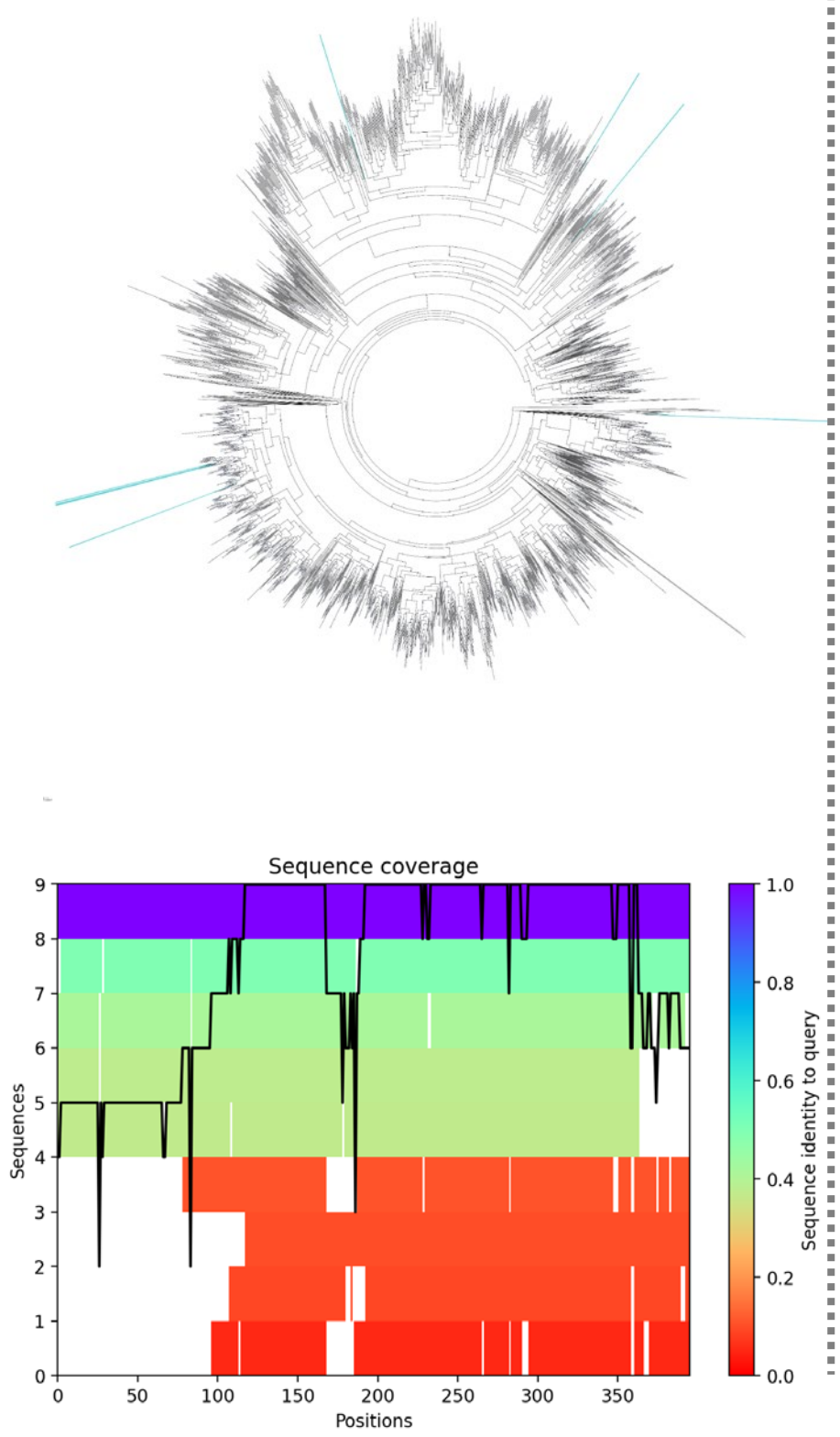

Purified sequences for inactive state (30th bin)

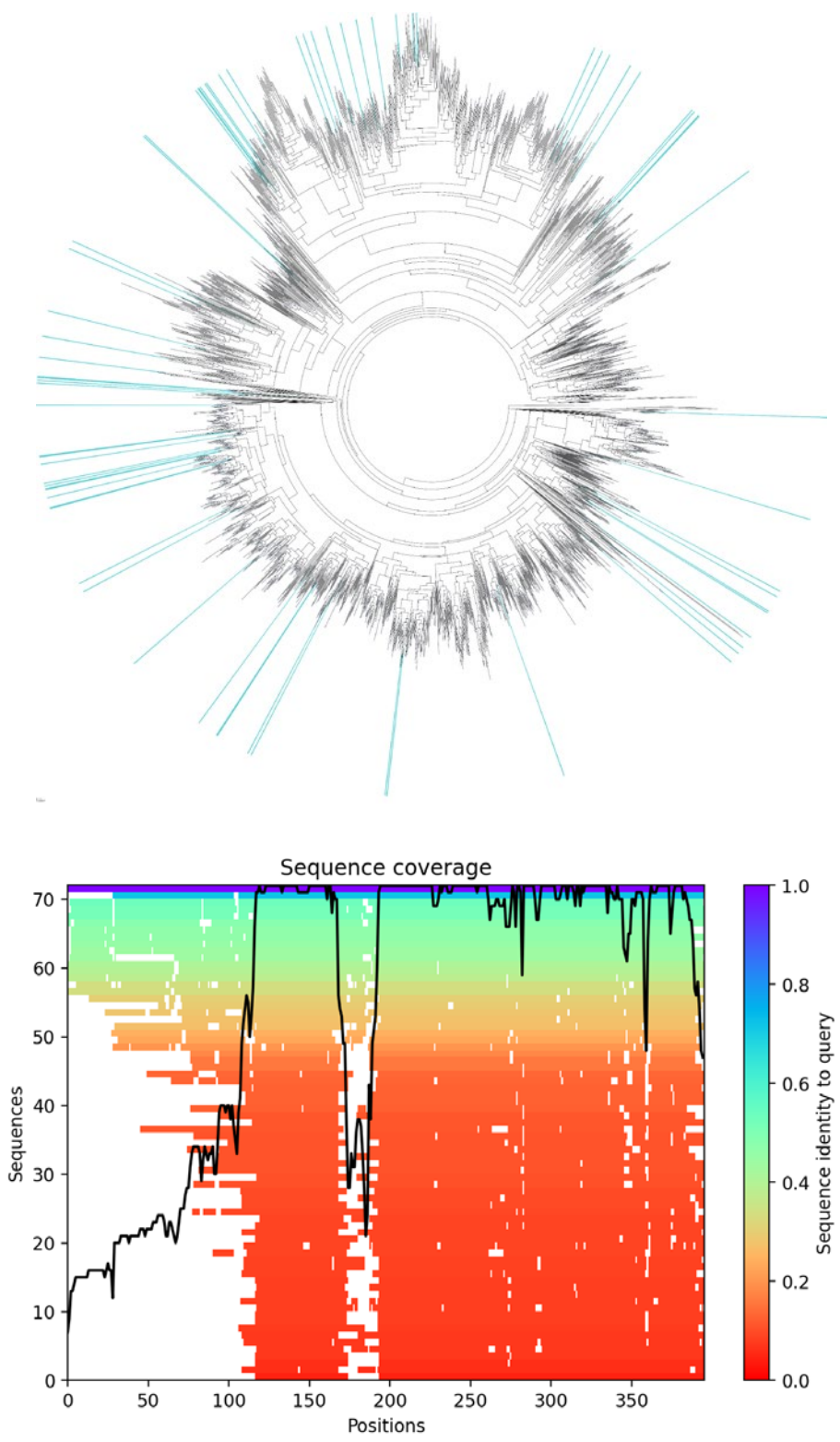

Purified sequences for inactive state (31st bin)

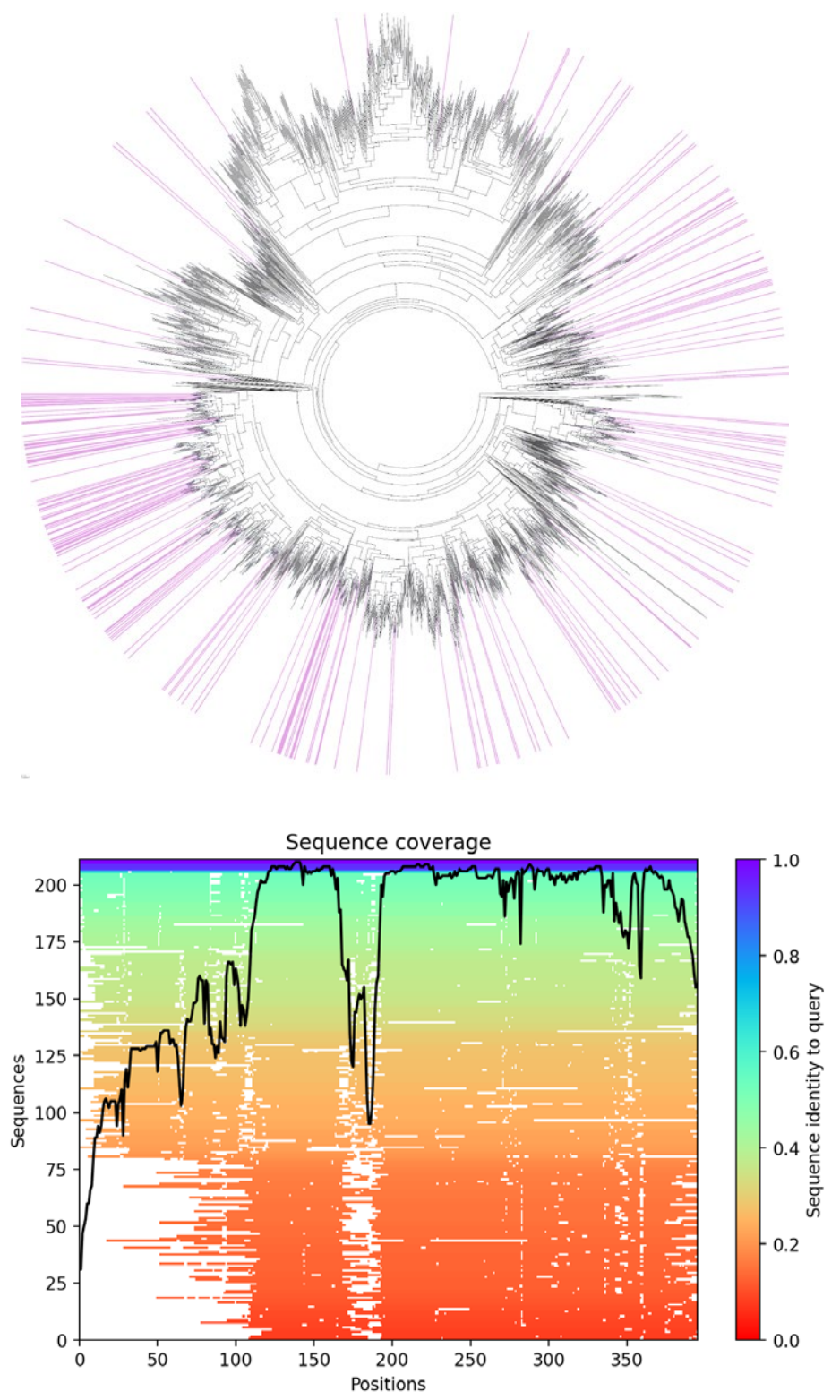

Purified sequences for active state (77th bin)

# Extended Data Figure. 10

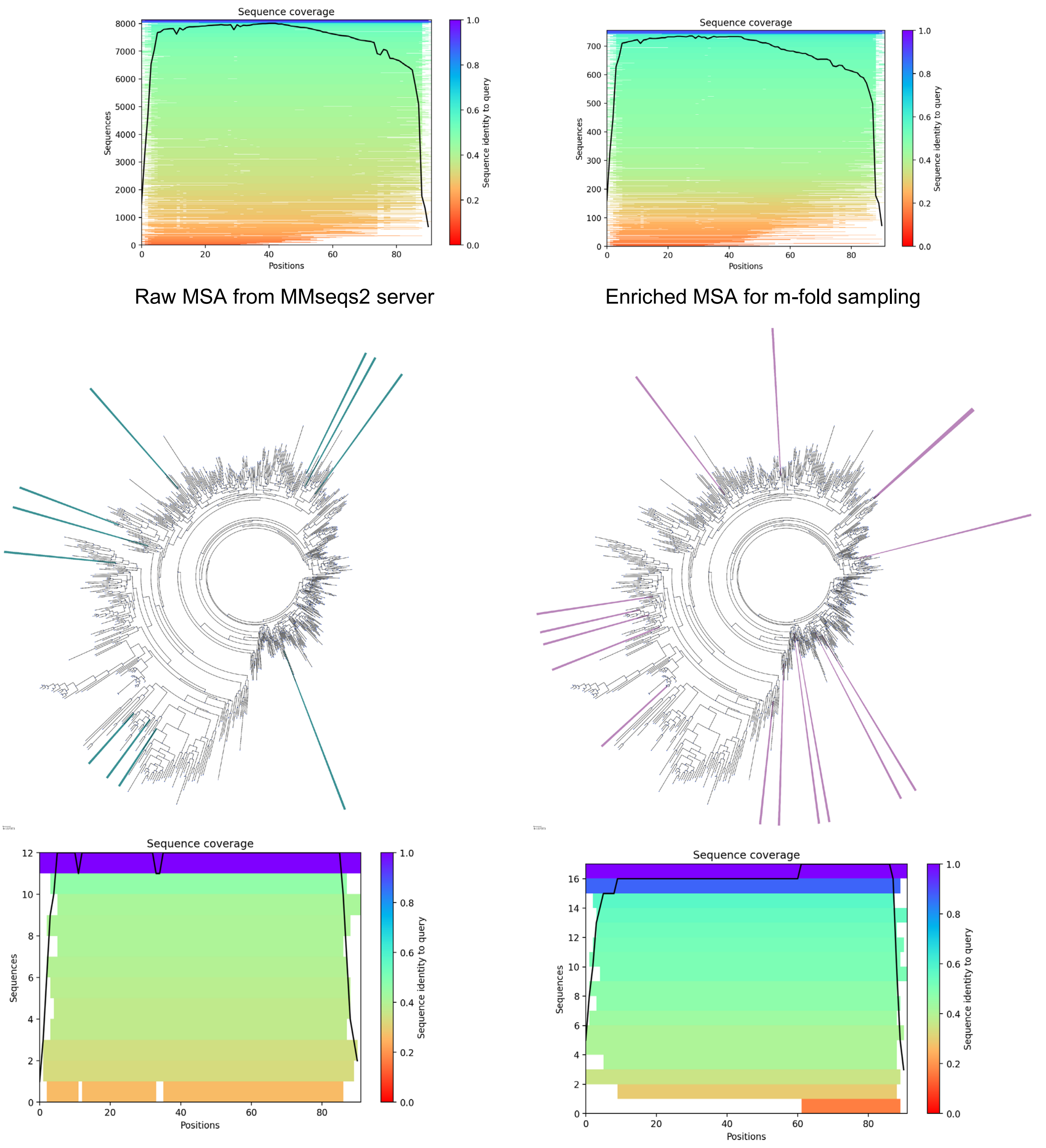

Purified sequences for fold-switch state (20th bin)

Purified sequences for ground state (26th bin)

# Extended Data Figure. 11

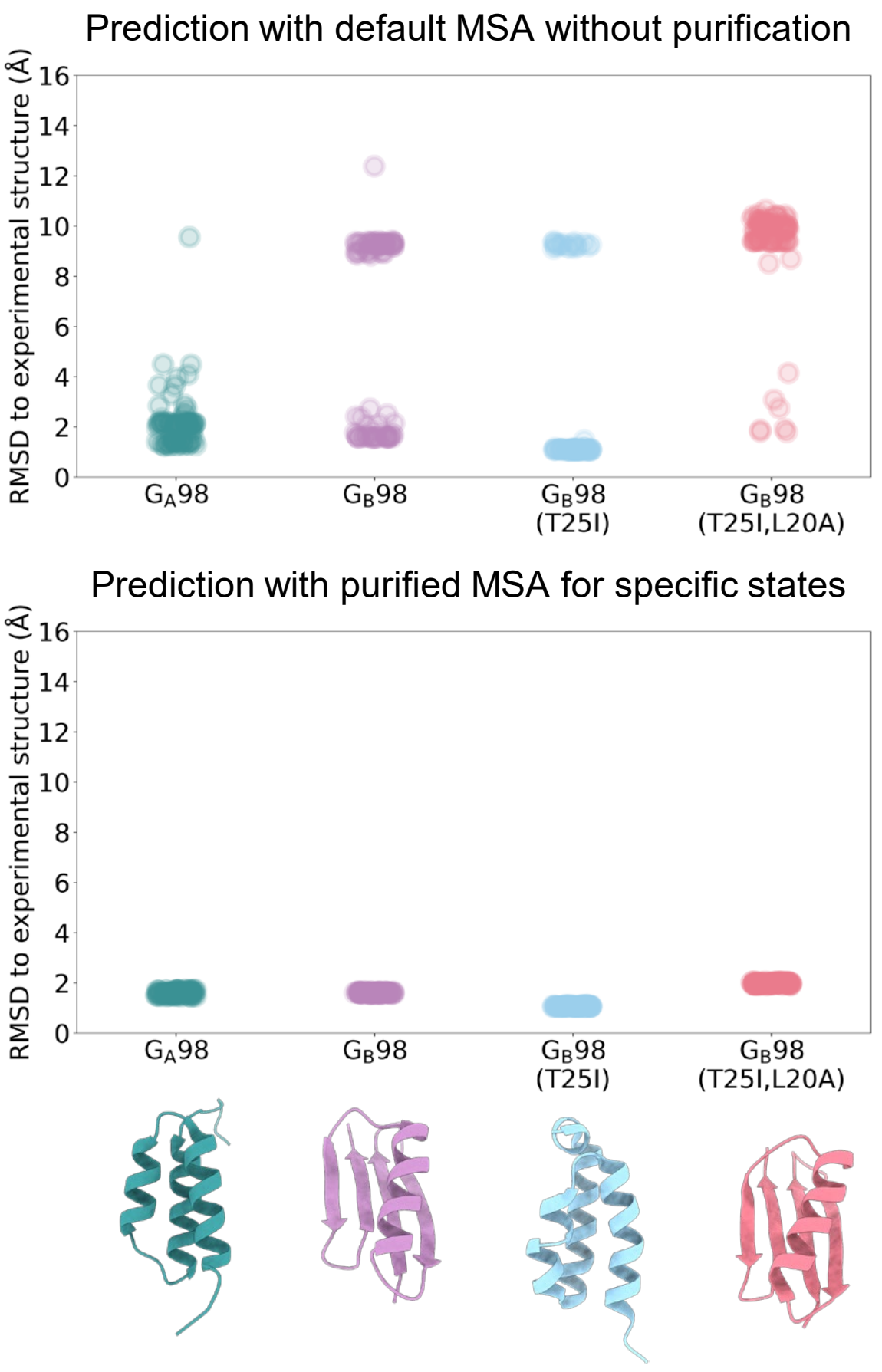

$4^{th}$ bin voted sequences for 3α conformation of $G_A98$:

```
--------LKNAKEDAIKELKEAGISQFFFNLINNAKTVEGVESLKNEI-------
--------LAEAKVLANRELDKYGVSDYYKNLINNAKTVEGVKALIDEI-------
------IKLSEAKAAAKAELKAAGVSDFFISKIDGAKTVEGVKALKELI-------
------IELAEAKDAALKEFDKYGVSDFYKKMIEKAGTVEGVKQLQSE--------
--------LKDAKEKAIEAIRKEGVSKLYEDLINKAKTIDGVNALRDQI-------
-------KLEFAKDAALKEFDKYGVSDFYKKMIEKAGTVEGVKQLQSE--------
-------NLSNAKKEAIEELKEAGTSDIYFDQINKANTIEGVEALKNET-------
-------NLSNAKKEAIGKLKEAGTSDIYFDQINKANTIEGVEALKNEI-------
--------LKNAKEDAIKELKEAGISQFFFNLINNAKTVEGVESLKNEI-------
--------LKDAKEKAIEAIRKEGVSKLYEDLINKAKTIDGVNALRDQI-------
```

$5^{th}$ bin voted sequences 3α conformation of $G_A98$:

```
TTYKLILNLKQAKEEAIKELVDAGIAEKYIKLIANAKTVEGVWTLKDEILTFTVTE
--------LAQAKEAAIKELKQYGIGDYYIKLINNAKTVEGVESLKNEILKALPTE
------------KEAVKAELKAAGASDFYTKKIDSANTVDGVKTLREMI-------
```

$5^{th}$ bin voted sequences for 4β1α conformation $G_B98$:

```
--YKLILNGKTLKGETTTEAVDAATAEKVFKQYANDNGVDGEWTYDDATKTFTVTE
-TYTLIIKGRTLTGTTTTKALSPEAAEKYFRNYATSNGIDTEWSYDKATRTFTVIE
--YYLVVNGQNAFYETLTKAVDAETARNAFIQSLKDDGVQGVWTYDDATKTFTV--
-TYHLVVNGKTLTATISYQATGTVQAGNYFENYVKSQGINADWSYDDVTRTFTVTE
-TYKLVVKTADTNITSEYDAANKEAAEEYFHAYVARNGLELNWTYDAETKTFMVT-
```